\documentclass[preprintnumbers,amsmath,amssymb,floatfix,11pt,prd,onecolumn,superscriptaddress,nofootinbib]{revtex4}
\usepackage[utf8]{inputenc}
\usepackage{diagbox}
\usepackage{latexsym}
\usepackage{epsfig}
\usepackage{epstopdf,float}
\usepackage{graphicx}
\usepackage{amssymb}
\usepackage{amsmath}
\usepackage{dcolumn}
\usepackage{bm}
\usepackage{color}
\usepackage{comment}
\usepackage[shortlabels]{enumitem}
\usepackage{subfigure}
\usepackage{multirow}
\usepackage{xcolor}
\begin{document}
\title{\bf Probing Non-rotating Black Hole in
Kalb-Ramond Gravity: Imaging and Polarized Signatures Surrounded by
Different Thick Accretion Flows}
\author{Xiao-Xiong Zeng}
\altaffiliation{xxzengphysics@163.com}\affiliation{College of
Physics and Electronic Engineering, Chongqing Normal University,
Chongqing $401331$, People's Republic of China}
\author{Chen-Yu Yang}
\altaffiliation{chenyu\_yang2024@163.com}\affiliation{Department of
Mechanics, Chongqing Jiaotong University, Chongqing $400000$,
People's Republic of China}
\author{Muhammad Israr Aslam}
\altaffiliation{mrisraraslam@gmail.com,
israr.aslam@umt.edu.pk}\affiliation{Department of Mathematics,
School of Science, University of Management and Technology,
Lahore-$54770$, Pakistan.}
\author{Rabia Saleem}
\altaffiliation{rabiasaleem@cuilahore.edu.pk}\affiliation{Department
of Mathematics, COMSATS University Islamabad, Lahore Campus,
Lahore-$54000$ Pakistan.}

\begin{abstract}
In this work, we consider a spherically symmetric static black hole
metric in Kalb-Ramond (KR) gravity, and investigate the impact of
relevant parameters on the black hole shadow and polarization
images. For black hole shadow images, we consider two geometrically
thick accretion disk models such as a phenomenological RIAF-like
model and an analytical HOU disk model. In each case, we observe a
bright ring-like structure corresponding to the higher-order images
with a surrounding region of non-zero intensity that represents the
primary image. The increasing values of $\hat{\lambda}$ or
$\hat{\gamma}$ results in decrease the size of the higher-order
image, while increasing values of observer inclination $\theta_{o}$
alter its shape and cause the horizons outline to be obscured. On
the other hand, in HOU disk model, at high observer inclinations,
the obscuration of the horizons outline by radiation from outside
the equatorial plane is weakened. Consequently, the brightness of
the primary image in the phenomenological model is significantly
greater than that in the HOU disk model, indicating the strong
gravitational lensing effect. For the polarized images, we use only
the HOU disk model with anisotropic radiation, assuming an infalling
accretion flow matter. The obtained results illustrate that the
polarization intensity $P_{o}$ in the higher-order image region is
significantly stronger than as compare to other regions, and it is
rapidly decreases away from this region. The variation in
$\hat{\lambda}$ and $\hat{\gamma}$ depicts the intrinsic structure
of the space-time and $\theta_o$ depends on the observers
orientation, together they shape the polarization features.
\end{abstract}
\date{\today}
\maketitle

\section{Introduction}
The expanding nature of our universe reflects its continuous
evolution, encompassing the vast array of experimental and
transformative processes that have occurred throughout time and
space, from the Big Bang to the emergence of humankind. The
exploration of various extended theories of gravity is primarily
driven by theoretical motivations, particularly the aim to address
inherent limitations and inconsistencies within general relativity
(GR). On the other hand, to account for recent observational
findings, dark matter and dark energy were considered, and to
develop a viable theory of gravity, modifications to GR at both
short and large scales are required \cite{th1}. With the detection
of gravitational waves, the recent observational era provides a new
window for testing gravitational theories in the strong-field
regime, and differentiating between them \cite{th2,th3,th4}.
Therefore, it is particularly important to gain intuition about the
properties of the compact objects predicted by different extended
theories, and the visual signatures, which they can introduce.
Consequently, to properly understanding the nature of gravity, it is
necessary that we consider hypothesis beyond the GR. To modify GR,
we need to alter the Einstein action and one way to do this is to
including the KR field \cite{th5}, which is interpreted as a
self-interacting second-rank antisymmetric tensor. The KR field may
be related with heterotic string theory \cite{th6} and can be
understood as stimulating closed strings. It has been observed that
spontaneous Lorentz invariance violation arises from the non-minimal
coupling between the gravitational sector and the non-zero vacuum
expectation value of the tensor field \cite{th7}. The KR field
interprets various properties, including deriving the third-rank
antisymmetric tensor. This tensor may be understood as a cause of
space-time torsion \cite{th8}.

The KR modification can also have an impact on the observed
anisotropy in the cosmic microwave background \cite{th9}. The KR
field has been widely studied in the framework of gravity and
particle physics \cite{th10,th11,th12,th13}. The striking similarity
between the KR field and space-time torsion indicates that GR, when
sourced by the KR field, is equivalent to a modified theory of
gravity incorporating torsion. Solar System tests of GR suggest that
the alterations in light bending or the perihelion precession of
Mercury caused by the KR field would be extremely small and remain
undetectable with current observational precision \cite{th14}. The
gravitational characteristics of the KR field and its impact on
particles have been investigated in \cite{th15}. In recent years,
the intriguing properties of the KR field have attracted
considerable attention, leading to numerous significant studies by
various authors. For example the authors in \cite{th16},
investigated the optical characteristics of rotating black hole in
the context of KR gravity. The Fermionic greybody factors and
quasinormal modes of black holes in the realm of KR gravity are
investigated in \cite{th17}. The parameter estimations as well as
some intricate properties of static and rotating KR black holes are
investigated in \cite{zubair}. The study of the optical and energy
processes around the black hole in the KR field for mass and
massless particles was discussed \cite{th18}. Recently, the research
community has launched marvellous studies on the implications of KR
gravity under effective implementations, including the gravitational
lensing of traversable wormholes \cite{th19}, bouncing cosmology
\cite{th20}, gravitational parity violation \cite{th21},
cosmological quantum entanglement \cite{th22} and many other
associated extensive studies are found in literature. Consequently,
in recent years, significant scholarly efforts have been devoted to
KR gravity, which potentially triggered the gravitational physics,
the beginning of a new channel to be understood as a manifestation
of a space-time enriched with distinctive geometrical properties.

Over the past decade, a major experimental breakthrough was achieved
with the detection of gravitational waves, opening a new avenue for
probing the nature of gravity \cite{sd1,sd2}. On the other hand, the
groundbreaking images of the supermassive black holes M$87^{\ast}$
and Sgr $A^{\ast}$ released by the Event Horizon Telescope (EHT)
have ushered in a new window of observational black hole
astrophysics \cite{sd3,sd4,sd6,sd10,jp20,jp21}. These images,
revealing distinctive dark shadows encircled by bright, asymmetric
emission rings, offer unprecedented opportunities to test GR in the
strong-field regime and to explore the intricate plasma environments
surrounding black holes \cite{th23}. The observed black hole shadow
encodes vital information about the space-time geometry, spin
orientation, and surrounding matter distribution of the black hole
\cite{th24}. Consequently, developing accurate theoretical models of
black hole shadows under realistic astrophysical conditions has
emerged as a leading frontier in gravitational physics. Black hole
imaging experiments, such as those conducted by the EHT, are deeply
rooted in the foundations of GR, which governs the motion of
particles and the propagation of photon in strong gravitational
field. These phenomena are governed by the theory of gravitational
lensing, which describes the observational effects arising from the
interaction of light with extreme space-time curvature.
Particularly, the intense gravitational field close to the black
holes gives rise to a distinct phenomenon known as the black hole
shadow. This effect occurs when a black hole is encircled by a
distribution of light sources, producing a dark silhouette against
the bright background.

The exploration of compact object shadows, particularly those of
black holes, has a long history and has now developed into a
relatively mature field of research. In this regard, an extensive
attention has been devoted to a variety of accretion models,
including spherical accretion models \cite{th25}, thin accretion
disk models \cite{gkref26,gkref27,sd37,sd38,zeng4}, and
geometrically thick disk models \cite{th26,th27,th28}. Subsequently,
significant progress have been made in the study of holographic
Einstein ring in the context of wave optics mechanism
\cite{sd21,sd22,sd23,israr1,israr22} as well as the shadow of
rotating black hole with celestial light sphere and thin accretion
disk \cite{sd34,sd35,sd36}. Among the various frameworks and
proposals to extend the study of black hole shadows through
different mechanisms, the polarized images exhibit the actual
astronomical environment around black holes. Recently, the EHT
unveiled polarized images of M$87^{\ast}$ and Sgr $A^{\ast}$
\cite{th29}. These observations reveal a remarkable polarization
pattern within the emission ring, where the linear polarization
images display a distinct spiral alignment of electric field
vectors. In black hole physics, polarization of radiation serves as
a crucial probe of magnetic field structures and the dynamics of the
surrounding matter. To extract polarization information from black
hole images, one must solve the null geodesic equations describing
photon trajectories in the black hole space-time. Additionally, the
parallel transport equations are solved to track the evolution of
polarization vectors along these geodesics \cite{th30}. Considering
the Schwarzschild black hole, the EHT employed the approximate
expression for null geodesics derived by Beloborodov \cite{th31} to
obtain the polarization images of a black hole surrounded by hot gas
\cite{th32}. This simplified model successfully replicates the
observed distribution of electric vector position angles and the
relative polarization intensity in the polarization image of
M$87^{\ast}$. In this scenario, the authors in \cite{th33} developed
a simplified model with an equatorial emission source, produced the
corresponding polarization images, and discussed the geometric
influence of black hole spin on photon parallel transport. These
studies reveal that the polarization properties are mainly governed
by the magnetic field geometry, with additional influence from the
black hole's involved parameters and the observer's position.
Moreover, the polarization images for various black hole models and
horizonless ultra-compact objects have also been discussed
extensively, for instance see Refs.
\cite{young1,th34,th35,th36,young2,th37,th38,th39,young3,zengnew1,zengnew2}.
Motivated by these groundbreaking studies, this work investigates
the influence of KR gravity on the shadow images under thick
accretion disks and polarization images of a static black hole.

The segments of this paper will show up in this order. In Sec. {\bf
II}, we will briefly explain about the background of the static
black hole in KR gravity as well as motion of photons around the
black hole. Section {\bf III} is dedicated to defining the electron
radiation, which including the radiative transfer equation and
synchrotron radiation. In the same section, we will also define a
comprehensive overview of accretion disk models, such as the
phenomenological model and the HOU disk model. In sect. {\bf IV}, we
will interpret the basic background of the polarized images with
anisotropic radiation. The visual characteristics of black hole
shadow as well as polarized images will be discussed in sect. {\bf
V}. Finally, the last section will be devoted to concluding remarks.
Throughout the manuscript, we consider geometric units with $G=c=1$,
where $c$ is the speed of light in vacuum and $G$ represent the
gravitational constant.

\section{Review of the non-rotating black hole in Kalb-Ramond gravity}
The action of the non-minimal coupling of the gravity in the
framework of KR field is defined as \cite{th7,lessa}
\begin{eqnarray}\nonumber
S&=&\int\sqrt{-g}d^4x\bigg[\frac{\hat{R}}{16\pi
G}-\frac{1}{12}H_{\varpi\xi\zeta}H^{\varpi\xi\zeta}-V\big(\bar{B}_{\xi\zeta}\bar{B}^{\xi\zeta}\pm
b_{\xi\zeta}b^{\xi\zeta}\big)+\frac{1}{16\pi
G}\big(\xi_2\bar{B}^{\xi\lambda}\bar{B}^\zeta_\lambda
R_{\xi\zeta}\\\label{pp1}&+&\xi_3\bar{B}_{\xi\zeta}\bar{B}^{\xi\zeta}\hat{R}\big)\bigg],
\end{eqnarray}
in which $g$ represent the determinant of the metric tensor and
$\hat{R}$ denotes the Ricci scalar. The tensor field
$\bar{B}_{\xi\zeta}$ illustrates the KR field that responsible the
Lorentz invariance violation satisfying
$\bar{B}_{\xi\zeta}=-\bar{B}_{\zeta\xi}$ with a non-zero expectation
value in vacuum $<\bar{B}_{\xi\zeta}>=b_{\xi\zeta}\neq0$ \cite{th5}.
With this expectation value, it is possible to decompose the tensor
field $\bar{B}_{\mu\nu}$ into a time-like vector and two space-like
vectors, similar to the decomposition of the electromagnetic field
tensor $F_{\xi\zeta}$ \cite{th7}. The KR field strength
$H_{\xi\zeta\gamma}=\partial_{\xi[}\bar{B}_{\zeta\gamma]}$ is itself
antisymmetric and admits an analogy with the electromagnetic field
tensor $F_{\xi\zeta}$, while $\bar{B}_{\xi\zeta}$ plays a role
analogous to the vector potential. Consequently, the KR action can
be constructed in a manner analogous to electrodynamics. The
potential $V$ is associated with the aforementioned vacuum
expectation value of $\bar{B}_{\xi\zeta}$, and $\xi_2$ and $\xi_3$
indicates the non-minimal coupling constants. Following
Ref.~\cite{lessa}, to study the effect of the KR vacuum expectation
value on the gravitational field, the KR field is taken in vacuum
such that
$\bar{B}_{\xi\zeta}\bar{B}^{\xi\zeta}=b_{\xi\zeta}b^{\xi\zeta}$. In
flat space-time, the Lorentz-violating expectation value
$b_{\xi\zeta}$ is constant, $\partial_\sigma b_{\xi\zeta}=0$,
admitting a constant norm
$j^2=\eta^{\varpi\upsilon}\eta^{\xi\zeta}b_{\varpi\xi}b_{\upsilon\zeta}$.
Under these conditions the KR field strength vanishes, yielding a
vanishing Hamiltonian \cite{th7}. Analogously, in curved space-time
the vacuum expectation value can be assumed covariantly constant,
$\nabla_\sigma b_{\xi\zeta}=0$, which again ensures both the
vanishing KR field strength and a vanishing Hamiltonian. Therefore,
when deriving black hole solutions in the background of KR gravity,
the KR vacuum expectation value $b_{\xi\zeta}$ is treated as a
constant tensor with constant norm and vanishing Hamiltonian. The
modified gravitational field equations can be written as
\begin{eqnarray}
\hat{R}_{\xi\zeta}-\frac{1}{2}\hat{R}g_{\xi\zeta}=\kappa
\bar{T}_{\xi\zeta}^{\mu_2}, \label{p2}
\end{eqnarray}
where $\hat{R}_{\mu\nu}$ is the Ricci tensor,
$\bar{T}_{\xi\zeta}^{\mu_2}$ is the energy-momentum tensor and the
four-dimensional static space-time in symmetric and spherical
geometry can defined as
\begin{eqnarray}
ds^2=-h(r)dt^2 +\frac{dr^2}{f(r)}+r^2\big(d\theta^2+\sin^2\theta
d\phi^2\big). \label{p3}
\end{eqnarray}
The ansatz for KR vacuum expectation value can be defined as
\begin{eqnarray}
b_2=-\hat{E}(r)dt\wedge dr, \label{p4}
\end{eqnarray}
in which $b_{tr}=-\hat{E}$. As defined earlier,
$b^2=g^{\varpi\upsilon}g^{\xi\zeta}b_{\varpi\xi}b_{\upsilon\zeta}$,
the norm of ansatz for KR vacuum expectation value is a constant
$b^2$ with metric (\ref{p3}) i.e.,
\begin{eqnarray}
\hat{E}(r)=|b|\sqrt{\frac{h(r)}{2f(r)}}, \label{p5}
\end{eqnarray}
in which $b$ indicates a constant. It is worthy to mention that the
function $\hat{E}(r)$ in Eq. (\ref{p5}) defines a static
pseudo-electric field in radial direction, such that
$\hat{E}^\xi=(0,\hat{E},0,0)$. Expanding the gravitational field
equations~(\ref{p2}) for the space-time metric~(\ref{p3}) and
solving for the metric functions, we find that $h(r) = f(r)$, such
that
\begin{equation}
h(r)=1-\frac{2M}{r}+\frac{\hat{\gamma}}{r^{\frac{2}{\hat{\lambda}}}}.
\label{p6}
\end{equation}
In this mechanism, we consider the spherically symmetric static
black hole metric in KR gravity, which can be written as
\begin{equation}
ds^2=-\left(1-\frac{2M}{r}+\frac{\hat{\gamma}}{r^{\tfrac{2}{\hat{\lambda}}}}\right)dt^2+\frac{dr^2}{1-\frac{2M}{r}+\frac{\hat{\gamma}}{r^{\tfrac{2}{\hat{\lambda}}}}}+r^2\big(d\theta^2+\sin^2\theta\,
d\phi^2\big),\label{eq:krm}
\end{equation}
where $\hat{\gamma}$ and $\hat{\lambda}$ are the associated
parameters of spontaneous Lorentz symmetry violation in KR gravity,
particularly, $\hat{\lambda}=|b|^2\xi_2$, i.e.,
$b^2=b_{\xi\zeta}b^{\xi\zeta}$, whereas $\hat{\gamma}$ is an
integration constant. Further, $M$ denotes the black hole mass, and
for convenience we set $M=1$ in the following calculations. The
event horizon of Eq. (\ref{eq:krm}) is determined by
\begin{equation}
\Delta=1-\frac{2M}{r}+\frac{\hat{\gamma}}{r^{\tfrac{2}{\hat{\lambda}}}}=0.
\end{equation}
For a finite value of $\hat{\gamma}$, the metric (\ref{eq:krm})
reduces to the Schwarzschild metric as $\hat{\lambda} \to 0$. When
$\hat{\lambda}=-1$ and $\hat{\lambda}=1$, the metric (\ref{eq:krm})
reduces to the Schwarzschild de-Sitter metric and the
Reissner-Nordstr\"{o}m metric, respectively. For $\hat{\lambda}>0$,
the metric (\ref{eq:krm}) is asymptotically flat, while for
$\hat{\lambda} \leq 0$, it becomes asymptotically non-flat. In this
work, we mainly focus on the asymptotically flat case, namely
$\hat{\lambda}>0$. For the derivation and detailed discussion of the
non-rotating black hole in KR gravity, see
Refs.~\cite{zubair,lessa,Xu2025}. The black hole shadow arises from
light emitted at infinity and observed by a distant observer. As
these light rays move towards the the black hole, the intense
gravitational field bends them toward the central dark region,
providing the background illumination of the shadow. Photons
traveling close to the position of the photon sphere, however,
primarily determine the shadow's structure. In this regard, to
determine the black hole shadow, it is essential to analyze the null
geodesic equations, which can be derived through the Hamilton-Jacobi
formalism \cite{carter}. In present work, we incorporate the
Lagrangian formalism, which are defined as follows
\begin{eqnarray}
\mathcal{L}(q,\dot{q})=\frac{1}{2}g_{\xi\zeta}\dot{q}^\xi\dot{q}^\zeta
\label{p8}
\end{eqnarray}
takes the form in the equatorial plane with $\theta=\frac{\pi}{2}$
and $\dot{\theta}=0$ as
\begin{eqnarray}
\mathcal{L}(q,\dot{q})=\frac{1}{2}\Big(-f(r)\dot{t}^2+\frac{\dot{r}^2}{f(r)}+r^2\dot{\phi}^2\Big),
\label{p9}
\end{eqnarray}
where ``dot'' is the derivative with respect to affine parameter
$\lambda$. Additionally, the photon motions satisfy the
Euler-Lagrange equation
\begin{eqnarray}
\frac{\partial\mathcal{L}}{\partial
q^\xi}-\frac{d}{d\lambda}\bigg(\frac{\partial\mathcal{L}}{\partial\dot{q}^\xi}\bigg)=0.
\label{p10}
\end{eqnarray}
Since the metric coefficients are independent of both time $t$ and
$\varphi$, so we have two conserved quantities, which are
$E=-f(r)dt/d\lambda$ and $L=r^{2}d\phi/d\lambda$, corresponds to the
energy and angular momentum of photons, respectively. The photon
motions satisfy the null geodesic equation, i.e., $\mathcal{L}=0$,
which leads that
\begin{eqnarray}
-f(r)\dot{t}^2+\frac{\dot{r}^2}{f(r)}+r^2\dot{\phi}^2=0, \label{p12}
\end{eqnarray}
and one can obtain the orbital equation for null geodesics as
\begin{eqnarray}
\frac{dr}{d\phi}=\pm
r\sqrt{f(r)}\sqrt{\hat{h}(r)^2\frac{E^2}{L^2}-1}, \label{p13}
\end{eqnarray}
with
\begin{eqnarray}
\hat{h}(r)^2=\frac{r^2}{f(r)}. \label{p14}
\end{eqnarray}
By applying the conditions
$\frac{dr}{d\phi}=0=\frac{d^2r}{d\phi^2}$, we obtain the necessary
relation for the radius of the circular null sphere as
\begin{eqnarray}
\frac{d}{dr}\Big(\hat{h}(r)^2\Big)\bigg|_{r=r_{ph}}=0, \label{p15}
\end{eqnarray}
in which $r_{ph}$ represent the radius of photon sphere. At the
equatorial plane of the black hole such that, $\theta = 90^\circ$,
the expression of the effective potential can be defined as
\begin{eqnarray}
\dot{r}^2+V_{eff}(r)=0 \label{p26}
\end{eqnarray}
where
\begin{eqnarray}
V_{eff}(r)=\frac{L^2f(r)}{r^2}-E^2. \label{p27}
\end{eqnarray}
Since, we are interested in the unstable null circular orbits close
to the black hole event horizon, where the photons revolve around
the black hole with constant radius, which is so-called position of
photon sphere. Photons that approaches the photon sphere suddenly
fall into the event horizon, whereas only those beyond the photon
sphere may approaches the observer. In this way, the radius of the
photon sphere, satisfies $\dot{r}=0$ and $\ddot{r}=0$. And the
constraints for a circular orbit is calculated by the effective
potential, which requires
\begin{equation}
    V_{eff}(r) = 0, \quad \frac{\partial V_{eff}(r)}{\partial r} = 0. \label{peq:effs}
\end{equation}
And for an unstable orbit satisfies the conditions, such as
$\frac{\partial^2 V_{eff}(r)}{\partial r^2} < 0$.

\section{Electron Radiation}

\subsection{\textbf{Radiative Transfer Equation}}

The covariant radiative transfer equation for an unpolarized light
is \cite{radiv}
\begin{equation}
    \frac{d}{d\lambda}\mathcal{I}=\mathcal{J}-\alpha\mathcal{I},\label{eq:rte}
\end{equation}
where $\mathcal{I},~\mathcal{J}$ and $\alpha$ are general
invariants, related to the physical quantities by
\begin{equation}
    \mathcal{I}=\frac{I_{\nu}}{\nu^{3}},\quad\mathcal{J}=\frac{j_{\nu}}{\nu^{2}},\quad\alpha=\nu\alpha_{\nu},\label{eq:rcs}
\end{equation}
with $I_{\nu}$ the specific intensity, $j_{\nu}$ the emissivity, and
$\alpha_{\nu}$ is the absorption coefficient. Hence, the solution of
Eq.~(\ref{eq:rte}) becomes
\begin{equation}
    \mathcal{I}(\lambda)=\mathcal{I}(\lambda_0)+\int_{\lambda_0}^\lambda d\lambda^{\prime}\mathcal{J}(\lambda^{\prime})\exp\left(-\int_{\lambda^{\prime}}^\lambda d\lambda^{\prime\prime}\alpha(\lambda^{\prime\prime})\right).
\end{equation}
To convert from the geometric unit system to the CGS system, one may
multiply the affine parameter in Eq.~(\ref{eq:rte}) by a factor
\begin{equation}
    \frac{ d}{ d\lambda}\to\frac{1}{\mathcal{C}}\frac{ d}{ d\lambda},\quad\mathcal{C}=\frac{r_{g}}{\nu_{0}},
\end{equation}
where $r_{g}=GM/c^{2}$ is the unit of length and $\nu_{0}$ is the
frequency of a real photon at infinity. And then Eq.~(\ref{eq:rte})
becomes
\begin{equation}
    \frac1{\mathcal{C}}\frac{ d}{ d\lambda}\mathcal{I}=\mathcal{J}-\alpha\mathcal{I},
\end{equation}
with the solution
\begin{equation}\label{sss8}
    I_{\nu}=g^{3}I_{\nu_{0}}+r_{g}\int_{\lambda_{0}}^{\lambda} d\lambda^{\prime}g^{2}j_{\nu}(\lambda^{\prime})\exp\left(-r_{g}\int_{\lambda^{\prime}}^{\lambda} d\lambda^{\prime\prime}\alpha_{\nu}(\lambda^{\prime\prime})/g\right),
\end{equation}
where $g=\nu_{0}/\nu$ is the redshift factor and $\nu$ is the photon
frequency in the local reference frame. If the four-velocity of the
fluid is $u^{\mu}$ and the local magnetic field is $B_{\mu}$
(satisfying $B_{\mu}$), then

\begin{eqnarray}\label{sss1}
g=\frac{k_{\mu}(\partial_{t})^{\mu}}{k_{\mu}u^{\mu}}=\frac{k_{t}}{k_{\mu}u^{\mu}}=\frac{-1}{k_{\mu}u^{\mu}}.
\end{eqnarray}
Hence, it is clear that the calculation of intensity requires both
the emissivity and the absorption coefficient. Moreover, in
Eq.~(\ref{eq:rcs}), the emission coefficient $j_{\nu}$ and the
absorption coefficient $\alpha_{\nu}$ depend on the radiation
process under consideration.

\subsection{\textbf{Synchrotron Radiation}}

In this work, the radiation process is taken to be synchrotron
emission of electrons in the extreme relativistic regime (in the CGS
unit system). In the following, $e$ denotes the elementary charge,
$c$ is the speed of light, $k_{B}$ is the Stefan Boltzmann constant,
and $h$ represents the Planck constant. In a plasma system,
synchrotron radiation is mainly contributed by electrons
\begin{equation}
    j_{\nu} = \frac{\sqrt{3} e^{3} B \sin{\theta_{B}}}{4 \pi m_{e} c^{2}} \int_{0}^{\infty} d\gamma \, N(\gamma) F\left( \frac{\nu}{\nu_{s}} \right),
\end{equation}
where $\gamma=\frac{1}{\sqrt{1-\beta^{2}}}$ is the Lorentz factor of
the charged particle, $N(\gamma)$ is the electron distribution
function, and $F(x)$ is related to the modified Bessel function of
the second kind $K_{n}(x)$
\begin{equation}
F(x)=x\int_{x}^{\infty}dy K_{5/3}(y).
\end{equation}
The angle $\theta_{B}$ is defined by the vectors $e^{\mu}_{(B)}$ and
$e^{\mu}_{(k)}$ as

\begin{equation}\label{sss2}
\theta_{B}=\arccos\bigg(e^{\mu}_{(B)}.e^{\mu}_{(k)}\bigg)=\arccos\bigg[\frac{g}{B}(B_{\mu
k^{\mu}})\bigg],
\end{equation}
with
\begin{equation}\label{sss3}
e^{\mu}_{(k)}=-\bigg(\frac{k^{\mu}}{u^{\nu}k}+u^{\mu}\bigg),
\end{equation}
and
\begin{equation}\label{sss4}
e^{\mu}_{(B)}=\frac{B^{\mu}}{B}.
\end{equation}
The characteristic frequency $\nu_{s}$ is given by
\begin{equation}\label{sss5}
\nu_{s}=\frac{3eB\sin\theta_{B}\gamma^{2}}{4\pi mc}.
\end{equation}
Different electron distributions correspond to different radiation
formulas. In this work, a thermal distribution is considered, having
the following distribution function, which is defined as
\begin{equation}
    N(\gamma) = \frac{n_{e} \gamma^{2} \beta}{\theta_{e} K_{2}(1/\theta_{e})} \exp\left( -\frac{\gamma}{\theta_{e}} \right),
\end{equation}
where $n_{e}$ is the electron number density, $\theta_{e} = k_{B}
T_{e} / m_{e} c^{2}$ is the dimensionless electron temperature, and
$T_{e}$ is the thermodynamic temperature of the electrons. In the
extreme relativistic case, where $\beta \approx 1$ and $\theta_{e}
\gg 1$, the asymptotic formula $K_{2}(1/\theta_{e}) \approx 2
\theta_{e}^{2}$ holds. Defining $z = \gamma / \theta_{e}$, we obtain
\begin{equation}
    j_{\nu} = \frac{\sqrt{3} n_{e} e^{3} B \sin{\theta_{B}}}{8 \pi m_{e} c^{2}} \int_{0}^{\infty} dz \, z^{2} \exp(-z) \, F\left( \frac{\nu}{\nu_{s}} \right).
\end{equation}
Letting $x = (\nu / \nu_{s}) z^{2}$, then emissivity becomes
\begin{equation}\label{pti1}
    j_{\nu} = \frac{n_{e} e^{2} \nu}{2 \sqrt{3} c \theta_{e}^{2}} I(x), \quad x = \frac{\nu}{\nu_{c}}, \quad \nu_{c} = \frac{3 e B \sin{\theta_{B}} \theta_{e}^{2}}{4 \pi m_{e} c},
\end{equation}
where the parameter $x$ is the ratio of the photon frequency $\nu$
to the characteristic frequency $\nu_{c}$ of the system. The
dimensionless function is defined as
\begin{equation}
    I(x) = \frac{1}{x} \int_{0}^{\infty} z^{2} \exp(-z) F\left( \frac{x}{ z^{2}} \right).
\end{equation}
The above expression does not have a closed analytic form and
requires the use of fitting functions depending on the specific
accretion disk model. In the context of a thermal electron
distribution, the absorption process follows Kirchhoff's law, which
states that all absorption coefficients must satisfy

\begin{eqnarray}\label{sss6}
\alpha_{\nu}=\frac{j_{\nu}}{B_{\nu}}, \quad
B_{\nu}=\frac{2h\nu^{3}}{c^{2}}\frac{1}{\exp(h\nu/k_{B}T_{e})-1},
\end{eqnarray}
where $B_{\nu}$ denotes the Planck blackbody radiation function.
Moreover, during the numerical simulations, the following constants
can be defined to simplify the computation
\begin{eqnarray}
C_{1}=\frac{n_{h}e^{2}\nu_{h}}{2\sqrt{3}c\theta^{2}_{h}},\quad
C_{2}=\frac{4\pi m_{e}c\nu_{h}}{3eB_{h}\theta^{2}_{h}}, \quad C_{3}=
\frac{h\nu_{h}}{m_{e}c^{2}}\frac{1}{\theta_{h}}, \quad
C_{4}=\frac{2h\nu^{3}_{h}}{c^{2}}, \quad
C_{5}=\sqrt{n_{h}m_{p}c^{2}},
\end{eqnarray}
in which $n_{h}$ and $\theta_{h}$ are evaluated at the event, with
$\nu_{h}=10^{9}$Hz$= 1$GHz and $B_{h}= 1$. Under this
parametrization,
\begin{eqnarray}\label{sss7}
j_{\nu}=C_{1}\bar{n_{e}}\bar{\nu}\tilde{\theta}^{-2}_{e}I(x), \quad
x=\frac{C_{2}\tilde{\nu}}{\tilde{B}\sin\theta_{B}\tilde{\theta}^{2}_{e}},\quad
B_{\nu}=\frac{C_{4}\tilde{\nu}^{3}}{\exp(C_{3}\tilde{\nu}/\bar{\theta}_{e})-1},
\end{eqnarray}
where
$\bar{\nu}=\frac{\nu}{\nu_{h}}$,~$\bar{n_{e}}=\frac{n_{e}}{n_{h}}$,
and $\bar{\theta}_{e}=\frac{\theta_{e}}{\theta_{h}}$. Based on
Eq.~(\ref{sss6}) and (\ref{sss7}), in principle one can determine
the intensity in Eq.~(\ref{sss8}). It should be noted, however, that
the electron number density, the electron temperature, and the
dimensionless function $I(x)$ are not yet specified. In the
following, we will discuss how to determine these parameters for
different accretion disk models.

\subsection{Accretion Disk Models}

In the unpolarized case, two geometrically thick and optically thin
accretion disk models can be considered, namely the phenomenological
model \cite{broderick2011evidence} and the HOU disk model
\cite{th27,hou2024new}.
\subsubsection{\textbf{Phenomenological Model}}

In cylindrical coordinates, we define $R=r\sin\theta$ as the
cylindrical radius and $z=r\cos\theta$ as the height above the
equatorial plane $\theta=\pi/2$. Following a construction method
similar to the analytical radiatively inefficient accretion flow
(RIAF) model \cite{broderick2011evidence}, the density and
temperature profiles can be defined as
\begin{equation}
    n_e = n_h\left(\frac r{r_h}\right)^2\exp\left(-\frac{z^2}{2(\alpha R)^2}\right),\quad T_e = T_h\left(\frac r{r_h}\right),
\end{equation}
where $n_h,\alpha,T_h$ are constants, and $r_h$ denotes the outer
horizon. Through the cold magnetization parameter
\begin{equation}
    \sigma=\frac{B^2}\rho=\frac{B^2}{n_e(m_pc^2)},
\end{equation}
the magnetic field strength can be defined as
\begin{equation}
B=\sqrt{\sigma\rho},
\end{equation}
where the dimensionless quantity $\rho=n_{e}(m_{p}c^{2})$ is the
fluid mass density. The parameter $\sigma$ may be set as a constant
or as a distribution \cite{pu2016effects}
\begin{equation}
    \sigma=\sigma_{h}\frac{r_{h}}{r},
\end{equation}
where $\sigma_{h}$ is the cold magnetization parameter at the
horizon. In this work, the typical order of magnitude is
$\sigma\sim0.1$. For the phenomenological model, two types of
electron radiation can be considered such as, isotropic radiation
and anisotropic radiation. For isotropic radiation, only the
strength of the magnetic field is taken into account, while the
direction is ignored, and thus the angle between the magnetic field
and the emitted photons is not considered, and the emissivity is
defined as
\begin{equation}
\bar{j}_{\nu}=\frac{1}{2}\int_{0}^{\pi}j_{\nu}\sin\theta_{B}d\theta_{B},
\end{equation}
with the corresponding fitting formula
\begin{eqnarray}
\bar{j}_{\nu}=\frac{ne^{2}\nu}{2\sqrt{3}c\theta^{2}_{e}}I(x), \quad
x=\frac{\nu}{\nu_{c}}, \quad \nu_{c}=\frac{3eB\theta^{2}_{e}}{4\pi
m_{e}c},
\end{eqnarray}
where the fitting function for $I(x)$ is given by
\cite{leung2011numerical}
\begin{equation}
    I(x)=\frac{4.0505}{x^{1/6}}\left(1+\frac{0.4}{x^{1/4}}+\frac{0.5316}{x^{1/2}}\right)\exp\left(-1.8899x^{1/3}\right).
\end{equation}
For anisotropic radiation, the direction of the magnetic field can
be modeled as a mixture of toroidal and poloidal components, with
the explicit form
\begin{equation}
B^\mu\sim(l,0,A,1),\label{eq:b}
\end{equation}
where $A$ is a tunable parameter, which is set to $0$ in this work.
The parameter $l$ is defined as
\begin{equation}
    l=-\frac{u_\phi}{u_t},\quad u_\nu=g_{\mu\nu}u^\mu=(u_t,u_r,u_\theta,u_\phi).
\end{equation}
This form of the magnetic field satisfies the orthogonality
condition with the fluid four-velocity, namely $u^{\mu}B_{\mu}=0$.
The emissivity corresponding to anisotropic radiation is given by
Eq.~(\ref{pti1}), and the fitting function for $I(x)$ is
\cite{mahadevan1996harmony}
\begin{equation}
I(x)=2.5651\left(1+1.92x^{-1/3}+0.9977x^{-2/3}\right)\exp\left(-1.8899x^{1/3}\right).\label{eq:ar}
\end{equation}
It is worth noting that, as mentioned above, for the
phenomenological model, the magnetic field direction does not need
to be considered in the case of isotropic radiation. In the case of
anisotropic radiation, the magnetic field direction is determined by
Eq.~(\ref{eq:b}). Therefore, polarization imaging is not taken into
account in either case. Finally, it is necessary to consider the
accretion flow motion pattern. The accreting matter can either be
assumed to move along geodesics or to follow a specified
four-velocity, with the latter generally including three different
models. In this scenario, we will considered three cases of motion
such as, orbiting motion, infalling motion and combined motion,
which are defined as below

\textbf{1. Orbiting motion} \cite{gold2020verification}

In this case the fluid rotates around the black hole, and the
four-velocity has only $u^{t}$ and $u^{\phi}$ components. The
explicit form of this motion is defined as
\begin{equation}
    u^\mu=u^t \{ 1, 0, 0, \Omega \},\label{eq:om}
\end{equation}
with
\begin{equation}
    u^t = \sqrt{-\frac{1}{g_{tt} + g_{\phi\phi} \Omega^2}}
    ,\quad\Omega = -\frac{g_{tt} l}{g_{\phi\phi}},\quad l =-\frac{u_\phi}{u_t}= l_0\frac{R^{3/2}}{\alpha + R},\quad R = r \sin\theta.
\end{equation}
where $l$ is the angular momentum density, and $l_{0}$ and $\alpha$
are tunable parameters chosen to ensure that the four-velocity
remains timelike throughout the spacetime. In this work we set
$l_0=\alpha=1$.

\textbf{2. Infalling motion}

Assuming that the fluid is at rest at infinity, i.e., $u_{t}=-1$,
the four-velocity is given by
\begin{equation}
    u^\mu=\{-g^{tt},-\sqrt{-(1+g^{tt})g^{rr}},0,0\}.\label{eq:im}
\end{equation}
The condition for the four-velocity to be timelike everywhere
requires $g^{tt}\leq-1$.

\textbf{3. Combined motion} \cite{pu2016effects}

The combined motion can be regarded as a mixture of the orbiting
motion and the infalling motion. For the orbiting motion part, we
have
\begin{equation}
    R = r \sin\theta,\quad l = \frac{R^{3/2}}{1 + R},\quad \Omega_o = -\frac{g_{tt}
    l}{g_{\phi\phi}},
\end{equation}
and for the infalling motion part, we have
\begin{equation}
    \Omega_f = 0,\quad u^r_f = -\sqrt{-(1 + g^{tt})g^{rr}}.
\end{equation}
Moreover, the combination of both motions is written as follows
\begin{equation}
    \Omega = \Omega_o + \beta_1 \left( \Omega_f - \Omega_o \right),\quad
    u^r = \beta_2 u^r_f,\quad
    u^t = \sqrt{ -\frac{ 1 + g_{rr} \left( u^r \right)^2 }
        { g_{tt} + g_{\phi\phi} \Omega^2 }}.
\end{equation}
Thus, the four-velocity is
\begin{equation}
    u^\mu=\{ u^t, u^r, 0, u^t\Omega \}.\label{eq:cm}
\end{equation}
For conveniently, in present work we set $\beta_1=\beta_2=0.2$, and
there is no other physical meanings.

\subsubsection{\textbf{HOU Disk Model}}

The HOU disk model is a steady-state axisymmetric accretion model
\cite{hou2024new,th27}. In this model, the accreting matter is
confined to surfaces of constant $\theta$, which leads that
$u^{\theta}\equiv0$. The conservation equation for mass flow then
reads
\begin{equation}
    \frac{d}{dr} \left( \sqrt{-g} \rho u^r \right) = 0,
\end{equation}
with the solution
\begin{equation}
    \rho = \rho_0 \frac{\sqrt{-g}u^r|_{r=r_0}}{\sqrt{-g}u^r},
\end{equation}
in which $\rho_0 = \rho(r_0)$ is the mass density at a reference
point. The projection of the conservation equation of the
energy-momentum tensor along $u^\mu$ is
\begin{equation}
    d e=\frac{e+p}\rho d \rho ,\label{eq:dedp}
\end{equation}
where $e$ is the internal energy of the fluid. Defining $k= T_p /
T_e$ as the proton-to-electron temperature ratio, the internal
energy of the fluid under this approximation satisfies
\begin{equation}
    e = \rho + \rho \frac{3}{2} (k+2) \frac{m_e}{m_p} \theta_e ,\label{eq:e}
\end{equation}
where $\theta_e = k_B T_e / m_e c^2$ is the dimensionless electron
temperature. Using the ideal gas equation of state, one obtains
\begin{equation}
    p = n k_B (T_p + T_e) = \rho (1+k) \frac{m_e}{m_p} \theta_e .\label{eq:p}
\end{equation}
Substituting Eqs.~(\ref{eq:e}) and (\ref{eq:p}) into
Eq.~(\ref{eq:dedp}) and integrating yields
\begin{equation}
    \theta_e = (\theta_e)_0 \left( \frac{\rho}{\rho_0}
    \right)^{\frac{2(1+k)}{3(2+k)}},
\end{equation}
where $(\theta_{e})_{0}=\theta_e(r_{0})$ is the temperature at a
reference point, typically chosen as $r_{0}=r_{h}$. From the
four-velocity of the fluid, the mass density and the dimensionless
electron temperature can be calculated as
\begin{equation}
    \rho(r,\theta) = \rho(r_h,\theta) \sqrt{\frac{R(r_h,\theta)}{R(r,\theta)}},
    \quad
    \theta_e(r,\theta) = \theta_e(r_h,\theta) \left[ \frac{R(r_h,\theta)}{R(r,\theta)} \right]^{\frac{1+k}{3(2+k)}},
\end{equation}
where $\rho(r_h,\theta)$ and $\theta_e(r_h,\theta)$ are the boundary
values at $r_h$. For convenience, we set $\rho(r_h,\theta)$ in the
$\theta$ direction as a Gaussian distribution, and in the conical
solution we take $\theta_e(r_h,\theta)$ to be a constant
\begin{equation}
    \rho(r_h,\theta) = \rho_h \exp \left[ - \left( \frac{\sin \theta - \sin \theta_J}{\sigma} \right)^2 \right],
    \quad
    \theta(r_h,\theta) = \theta_h ,
\end{equation}
where $\theta_J$ is the average position in the $\theta$ direction,
and $\sigma$ describes the standard deviation of the distribution.
For M87*, observational results indicate
$\rho_{h}\approx1.5\times10^{3}\,\mathrm{g/cm/s^{2}}$ and
$\theta_{h}\approx16.86$, corresponding to $n_h = 10^6
\,\mathrm{cm^{-3}}$ and $T_h = 10^{11} \,\mathrm{K}$ \cite{umt1}.
For a spherically symmetric space-time, the magnetic field
configuration simplifies to
\begin{equation}
    B^{\mu}=\frac{\Psi}{\sqrt{-g}\,u^{r}}\left(u_{t}u^{\mu}+\delta_{t}^{\mu}\right),
\end{equation}
where $\Psi = F_{\theta\phi}$ is a component of the electromagnetic
tensor. Noting that $u^r$ appears in the denominator, the fluid in
the HOU disk model cannot be take the orbiting motion. We adopt the
split-monopole solution, which are defined as follows
\begin{equation}
    \Psi = \Psi_0 \,\mathrm{sign}(\cos \theta)\,\sin \theta .
\end{equation}
This analysis indicates that the HOU disk requires consideration of
the magnetic-field direction, and therefore the electron radiation
model is taken to be anisotropic radiation, as defined in
Eq.~(\ref{eq:ar}). For the HOU disk model, the four-velocity of the
fluid can be chosen as the infalling motion (see Eq. (\ref{eq:im}))
or the combined motion (see (\ref{eq:cm})) as previously defined.
One may also consider the ballistic approximation, in which the
fluid moves along geodesics (for related calculations in Kerr
space-time, see Ref.~\cite{hou2024new}). However, the orbiting
motion as defined in Eq. (\ref{eq:om}) cannot be adopted, since in
the HOU disk model the fluid requires a radial velocity; otherwise,
the magnetic field would diverge. The function $I(x)$ takes the
anisotropic synchrotron radiation fitting form given in
Eq.~(\ref{eq:ar}).

\section{Polarized Imaging}
For polarized imaging, only the HOU disk model with anisotropic
radiation needs to be considered. As defined in the previous
section, the accretion flow motion mode is chosen to be the
infalling motion. In the WKB approximation, the propagation of light
rays satisfies the radiative transfer equation
\begin{equation}
    k^\mu \nabla_\mu S^{\alpha\beta} = J^{\alpha\beta} + H^{\alpha\beta\mu\nu} S_{\mu\nu},\label{eq:rte2}
\end{equation}
where $k^\mu$ is the wave vector of the light ray, $S^{\alpha\beta}$
is the polarization tensor describing the polarization state of the
light, $J^{\alpha\beta}$ characterizes the emission properties of
the radiation source, and $H^{\alpha\beta\mu\nu}$ represents the
response of the medium to the light propagation, usually including
absorption and Faraday rotation effects. For numerical solutions of
the radiative transfer equation, one may refer to the open-source
code Coport 1.0 \cite{huang2024coport}. Building upon Coport 1.0, we
can make use of the gauge invariance of $S^{\alpha\beta}$ to
simplify the computation in a simple parallel-transported tetrad. In
this case, the covariant radiative transfer, as defined in Eq.
(\ref{eq:rte2}) is decomposed into two parts. The first part
reflects the gravitational effect,
\begin{equation}
    k^\mu \nabla_\mu f^a = 0, \quad f^a k_a^{} = 0 ,
\end{equation}
where $f^\mu$ is a normalized spacelike vector orthogonal to
$k^\mu$. The second part is the radiative transfer equation,
\begin{equation}
    \frac{d}{d\lambda} S = R(\chi)J - R(\chi)MR(-\chi)S ,
\end{equation}
where
\begin{equation}
    S=\begin{pmatrix}\mathcal{I}\\Q\\\mathcal{U}\\\mathcal{V}\end{pmatrix},\quad J=\frac{1}{\nu^2}\begin{pmatrix}j_I\\j_Q\\j_U\\j_V\end{pmatrix},\quad M=\nu\begin{pmatrix}a_I&a_Q&a_U&a_V\\a_Q&a_I&r_V&-r_U\\a_U&-r_V&a_I&r_Q\\a_V&r_U&-r_Q&a_I\end{pmatrix}.
\end{equation}
$R(\chi)$ is the rotation matrix, which accounts for the rotation
between the synchrotron emission basis and the parallel-transported
reference basis
\begin{equation}
    R(\chi) =
    \begin{pmatrix}
        1 &  &  &  \\
        & \cos (2\chi) & -\sin (2\chi) &  \\
        & \sin (2\chi) & \cos (2\chi) &  \\
        &  &  & 1
    \end{pmatrix} .\label{eq:rx}
\end{equation}
The rotation angle $\chi$ is the angle between the reference vector
$f^\mu$ and the local magnetic field $b^\mu$ in the transverse
subspace of the light ray
\begin{equation}
    \chi = \mathrm{sign}(\epsilon_{\mu\nu\alpha\beta} u^\mu f^\nu b^\rho k^\sigma)
    \arccos \left( \frac{P^{\mu\nu} f_\mu b_\nu}{\sqrt{(P^{\mu\nu} f_\mu f_\nu)(P^{\alpha\beta} b_\alpha b_\beta)}} \right) ,
\end{equation}
where $P^{\mu\nu}$ is the induced metric in the transverse subspace.
At the observer, the Stokes parameters need to be projected onto the
observer’s screen with the aid of the rotation matrix
(\ref{eq:rx}), where the rotation angle is
\begin{equation}
    \chi_{o}=\mathrm{sign}(\epsilon_{\mu\nu\rho\sigma}u^{\mu}f^{\nu}d^{\rho}k^{\sigma})\arccos\left(\frac{P^{\mu\nu}f_{\mu}d_{\nu}}{\sqrt{(P^{\mu\nu}f_{\mu}f_{\nu})\left(P^{\alpha\beta}d_{\alpha}d_{\beta}\right)}}\right),
\end{equation}
with $d^\mu$ being the $y$-axis direction of the screen. In this
work we choose $d^{\mu}=-\partial_{\theta}^{\mu}$. The projection
results are
\begin{equation}
    \mathcal{I}_{o}=\mathcal{I},\quad\mathcal{Q}_{o}=\mathcal{Q}\cos\chi_{o}-\mathcal{U}\sin\chi_{o},\quad\mathcal{U}_{o}=\mathcal{Q}\sin\chi_{o}+\mathcal{U}\cos\chi_{o},\quad\mathcal{V}_{o}=\mathcal{V}.
\end{equation}

\begin{figure}[H]
\centering \subfigure{\includegraphics[scale=0.5]{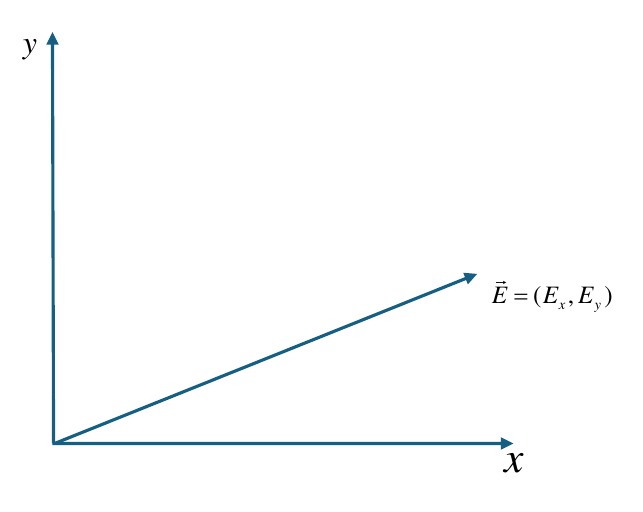}}
\caption{For the electric field vector $\vec{E}$ shown in the
figure, $\mathcal{Q}_{o}>0$ and $\mathcal{U}_{o}>0$.} \label{fig1}
\end{figure}

The Stokes parameter $\mathcal{I}_{o}$ represents the intensity of
the radiation. As $\mathcal{V}_{o}$ encodes the information of
polarization: if $\mathcal{V}_{o}$ is positive, the radiation is
left-handed polarization; otherwise, it is right-handed
polarization. $\mathcal{Q}_{o}$ and $\mathcal{U}_{o}$ together
describe the information of the electric field $\vec{E}=(E_x,E_y)$
as
\begin{equation}
    \mathcal{Q}_{o}=E_x^2-E_y^2,\quad \mathcal{U}_{o}=2E_x E_y.
\end{equation}
If $\mathcal{U}_{o}$ is positive, $E_x$ and $E_y$ have the same
sign, and $\vec{E}$ lies in the first and third quadrants; if
$\mathcal{U}_{o}$ is negative, $E_x$ and $E_y$ have opposite signs,
and $\vec{E}$ lies in the second and fourth quadrants. The sign of
$\mathcal{Q}_{o}$ reflects whether $\vec{E}$ is aligned with the
line $y=x$ or $y=-x$. On the projection screen, the polarization
vector $\vec{f}$ corresponds to the polarization intensity and the
electric vector position angle (EVPA), which are given by
\begin{equation}
    P_{o}=\sqrt{\mathcal{Q}_{o}^{2}+\mathcal{U}_{o}^{2}},\quad \Phi_{\mathrm{EVPA}}=\frac{1}{2}\arctan\frac{\mathcal{U}_{o}}{\mathcal{Q}_{o}},\label{eq:pv}
\end{equation}
and the corresponding schematic diagram are presented in Fig.
\textbf{\ref{fig1}}.

\section{Numerical Results}
In order to observe the black hole shadow and polarization images of
non-rotating KR black hole on the observer's screen, we consider a
zero angular momentum observer (ZAMO). Particularly, the ZAMO is
positioned at coordinates ($t_{o},~r_{o},~\theta_{o},~\phi_{o}$) and
a locally orthogonal normalized frame exists in the neighborhood of
the observer, which can be define as
\begin{eqnarray}\nonumber
&&e_{(t)}=\bigg(\sqrt{-\frac{1}{g_{tt}}},~0,~0,~0\bigg),\;\;\;\;\;\;\;\;\;\;\;
e_{(r)}=\bigg(-\sqrt{\frac{1}{g_{rr}}},~0,~0,~0\bigg)\\\label{s7}&&
e_{(\theta)}=\bigg(0,~\sqrt{\frac{1}{g_{\theta\theta}}},~0,~0\bigg),\;\;\;\;\;\;\;\;\;\;\;
e_{(\phi)}=\bigg(-\sqrt{\frac{1}{g_{\phi\phi}}},~0,~0,~0,\bigg),
\end{eqnarray}
Within the framework of the ZAMO, the $4$-momentum of photons can be
illustrated as $p_{(\xi)}=p_{\zeta}e^{\zeta}_{(\xi)}$, where
$p_{(\xi)}$ and $p_{(\zeta)}$ represent the four-momentum in the
ZAMO frame and the Boyer-Lindquist coordinate system, respectively.
The relationship between photon $4$-momentum and the celestial
coordinates $(\hat{\mu},~\hat{\nu})$ are defined as \cite{sd35}
\begin{eqnarray}\label{s9}
\cos\hat{\mu}=\frac{p^{(r)}}{p^{(t)}},\quad
\tan\hat{\nu}=\frac{p^{(\phi)}}{p^{(\theta)}}.
\end{eqnarray}
The observer frame can be equipped with a standard Cartesian
coordinate system $(\hat{\alpha},~\hat{\beta})$, which developed an
accurate align with celestial coordinates as
\begin{eqnarray}\label{s10}
\hat{\alpha}=-2\tan\frac{\hat{\mu}}{2}\sin\hat{\nu}, \quad
\hat{\beta}=-2\tan\frac{\hat{\mu}}{2}\cos\hat{\nu}.
\end{eqnarray}

\subsection{Phenomenological Model with Isotropic Radiation}

\begin{figure}[H]
\centering
\subfigure[$\hat{\gamma}=0.001,\theta_{o}=0.001^\circ$]{\includegraphics[scale=0.4]{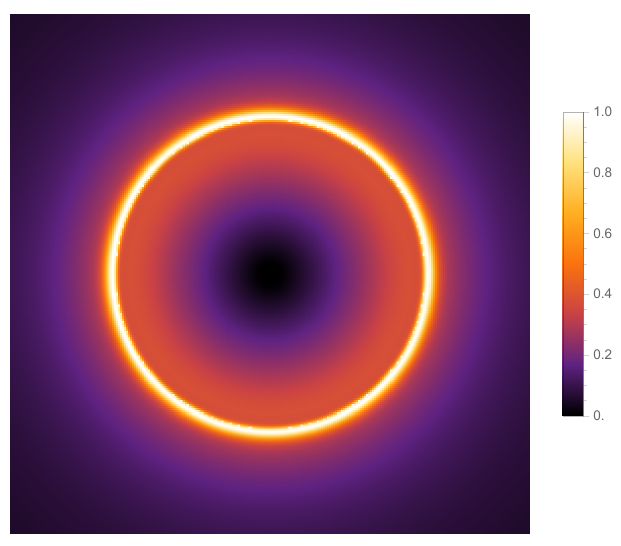}}
\subfigure[$\hat{\gamma}=0.001,\theta_{o}=17^\circ$]{\includegraphics[scale=0.4]{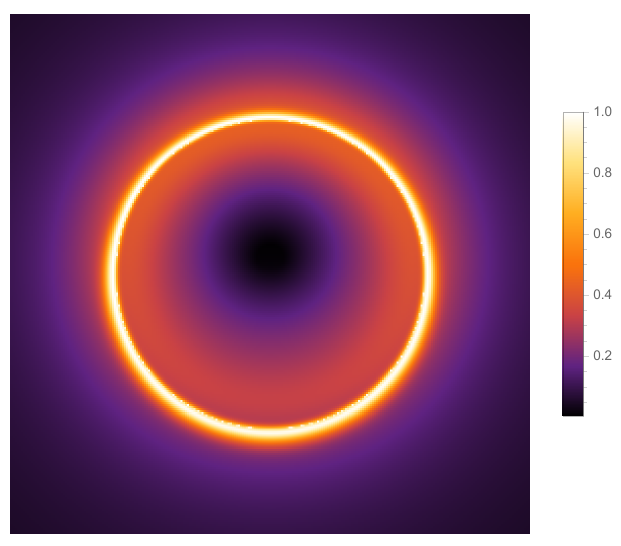}}
\subfigure[$\hat{\gamma}=0.001,\theta_{o}=75^\circ$]{\includegraphics[scale=0.4]{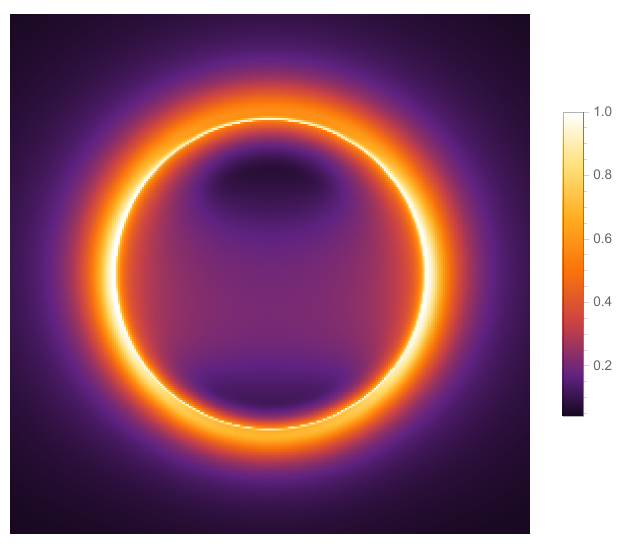}}
\subfigure[$\hat{\gamma}=0.55,\theta_{o}=0.001^\circ$]{\includegraphics[scale=0.4]{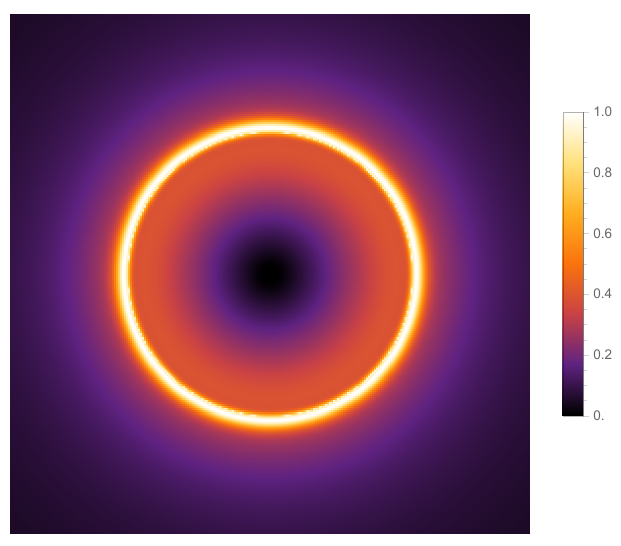}}
\subfigure[$\hat{\gamma}=0.55,\theta_{o}=17^\circ$]{\includegraphics[scale=0.4]{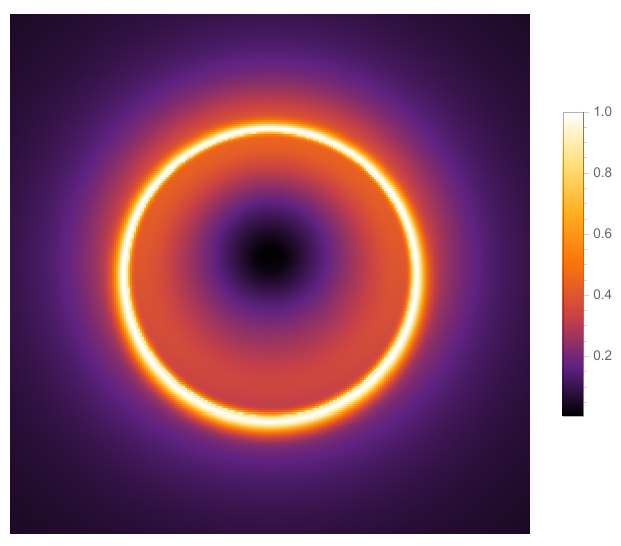}}
\subfigure[$\hat{\gamma}=0.55,\theta_{o}=75^\circ$]{\includegraphics[scale=0.4]{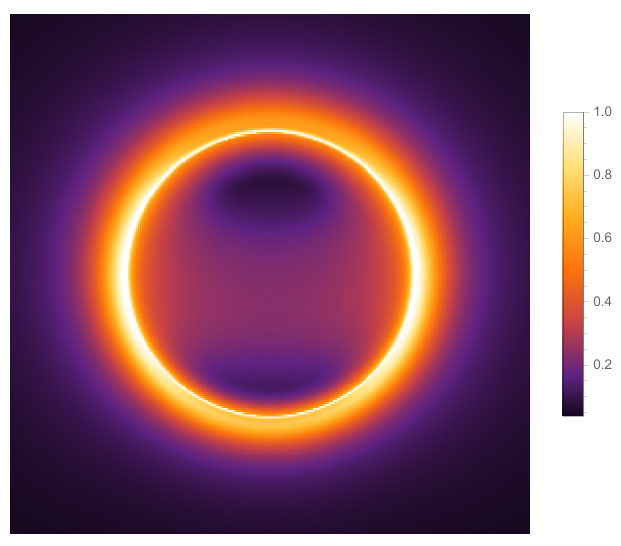}}
\subfigure[$\hat{\gamma}=0.95,\theta_{o}=0.001^\circ$]{\includegraphics[scale=0.4]{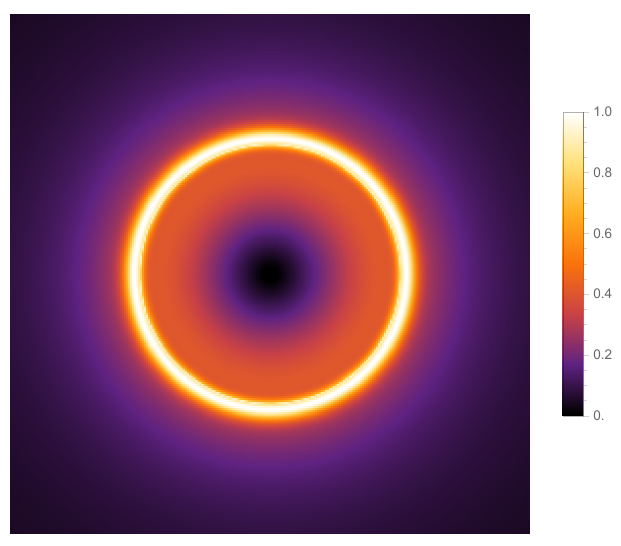}}
\subfigure[$\hat{\gamma}=0.95,\theta_{o}=17^\circ$]{\includegraphics[scale=0.4]{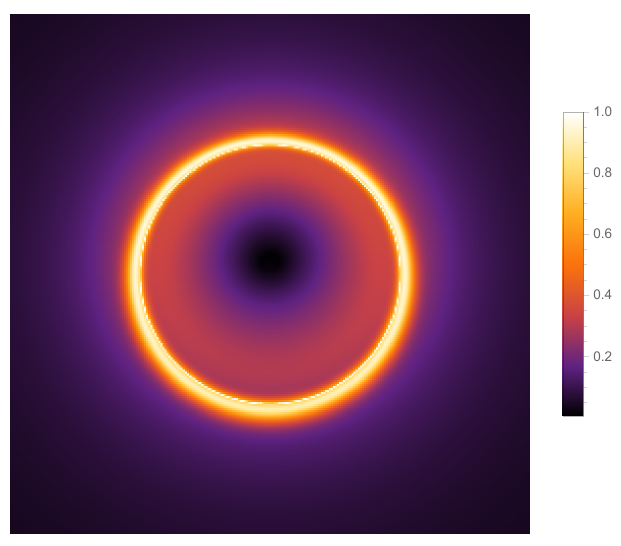}}
\subfigure[$\hat{\gamma}=0.95,\theta_{o}=75^\circ$]{\includegraphics[scale=0.4]{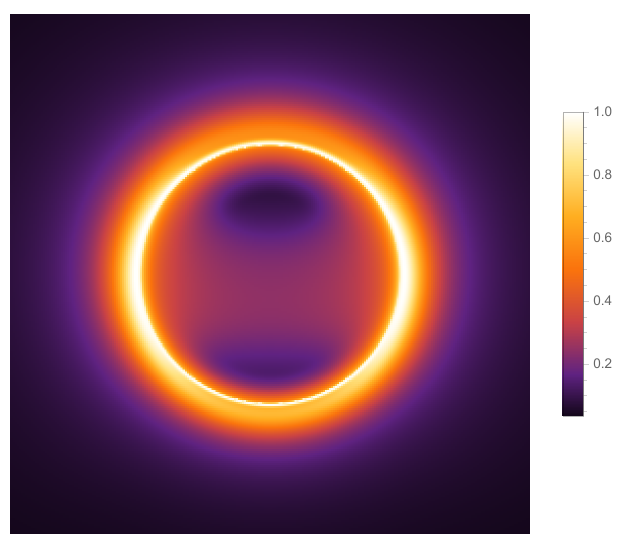}}
\caption{Black hole shadow images of the phenomenological model with
isotropic radiation. The accretion flow motion mode is the infalling
motion. The parameters are fixed at $\hat{\lambda}=0.9$ and the
observing frequency is $230\,\mathrm{GHz}$.}\label{fig2}
\end{figure}
\begin{figure}[H]
\centering
\subfigure[$\hat{\lambda}=0.1,\theta_{o}=0.001^\circ$]{\includegraphics[scale=0.4]{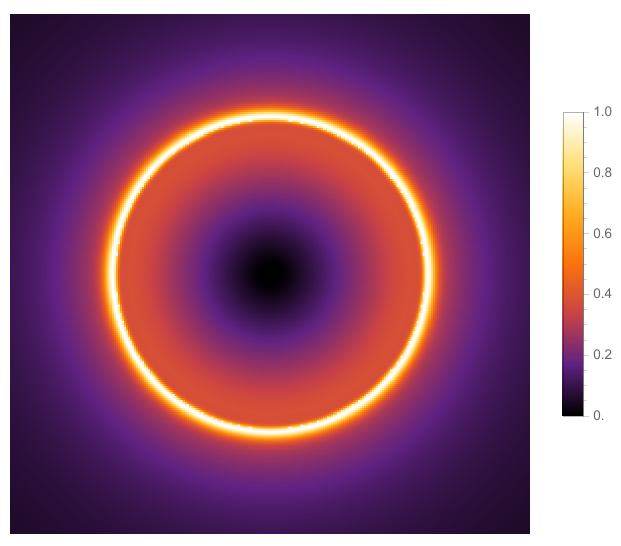}}
\subfigure[$\hat{\lambda}=0.1,\theta_{o}=17^\circ$]{\includegraphics[scale=0.4]{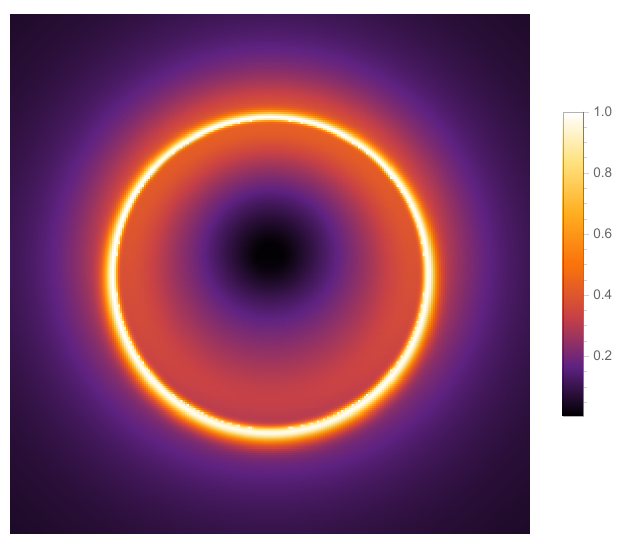}}
\subfigure[$\hat{\lambda}=0.1,\theta_{o}=75^\circ$]{\includegraphics[scale=0.4]{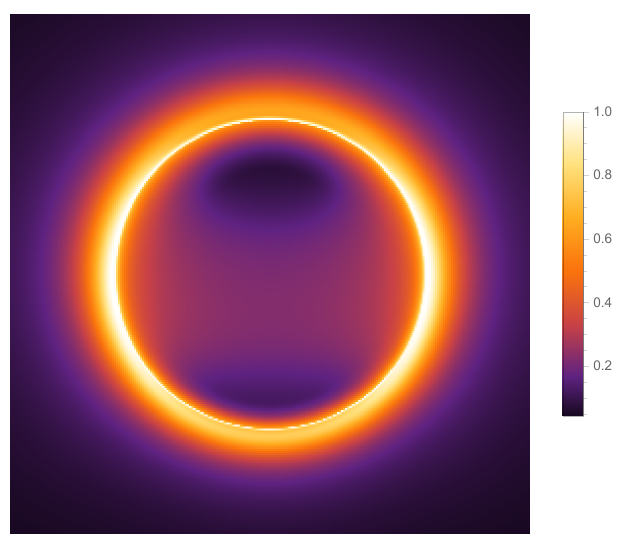}}
\subfigure[$\hat{\lambda}=0.55,\theta_{o}=0.001^\circ$]{\includegraphics[scale=0.4]{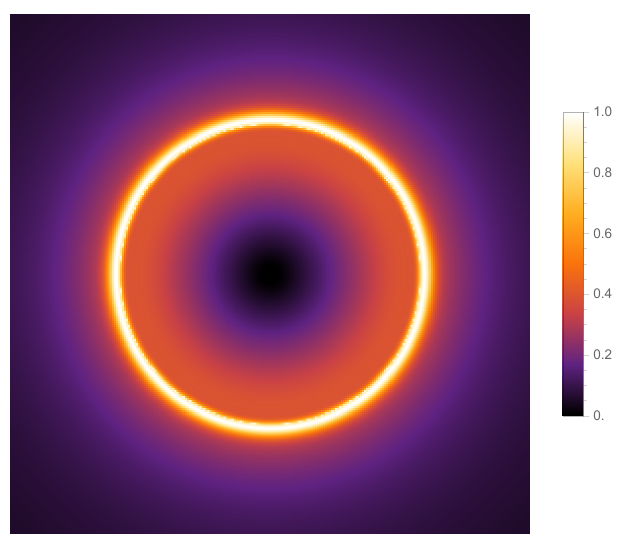}}
\subfigure[$\hat{\lambda}=0.55,\theta_{o}=17^\circ$]{\includegraphics[scale=0.4]{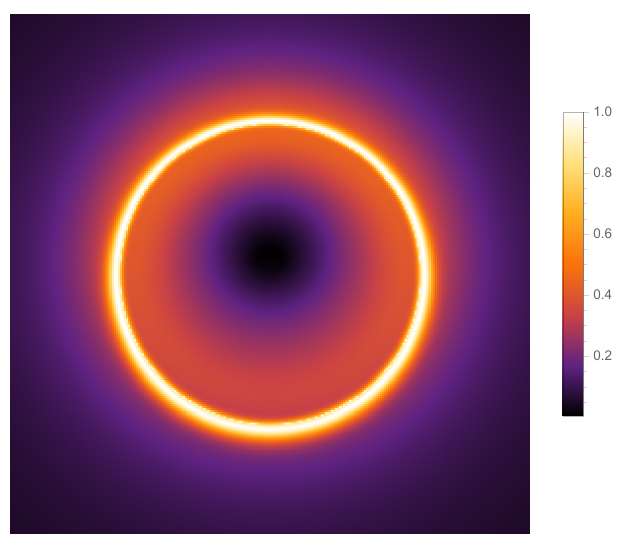}}
\subfigure[$\hat{\lambda}=0.55,\theta_{o}=75^\circ$]{\includegraphics[scale=0.4]{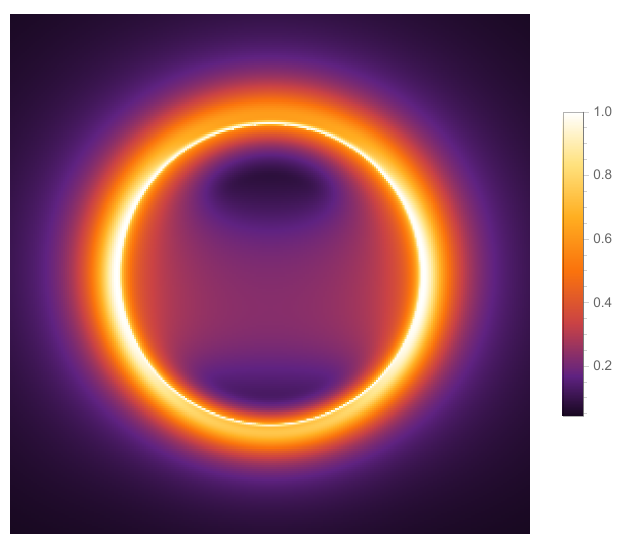}}
\subfigure[$\hat{\lambda}=0.9,\theta_{o}=0.001^\circ$]{\includegraphics[scale=0.4]{Iso7.pdf}}
\subfigure[$\hat{\lambda}=0.9,\theta_{o}=17^\circ$]{\includegraphics[scale=0.4]{Iso8.pdf}}
\subfigure[$\hat{\lambda}=0.9,\theta_{o}=75^\circ$]{\includegraphics[scale=0.4]{Iso9.pdf}}
\caption{Black hole shadow images of the phenomenological model with
isotropic radiation. The accretion flow motion mode is the infalling
motion. The parameters are fixed at $\hat{\gamma}=0.95$ and the
observing frequency is $230\,\mathrm{GHz}$.} \label{fig3}
\end{figure}

Figures \textbf{\ref{fig2}} and \textbf{\ref{fig3}} shows that the
black hole shadow images of the phenomenological model with
isotropic radiation. The observing frequency is fixed at $230
\,\mathrm{GHz}$, and the accretion flow motion mode is chosen to be
the infalling motion. In Fig.~\textbf{\ref{fig2}}, the associated
parameter of spontaneous Lorentz symmetry violation is fixed at
$\hat{\lambda}=0.9$, while the parameter $\hat{\gamma}$ is taken as
$0.001, 0.05, 0.95$. The observer inclination angles are chosen as
$\theta_o=0.001^\circ,~17^\circ,~75^\circ$. From
Fig.~\textbf{\ref{fig2}} it can be seen that in all images a bright
ring-like structure appears, corresponding to higher-order images,
i.e., photons that orbit the black hole once or multiple times
before reaching the observer. Outside this ring-like structure,
there exists a region with nonzero intensity, corresponding to the
primary image, where photons travel directly from the accretion disk
to the observer. It is worth noting that for all parameters, inside
the higher-order image there exists a region with zero intensity,
which originates from the event horizon of the black hole. For
geometrically thin accretion disks, this region is referred to as
the inner shadow, which may potentially be captured by the EHT
\cite{chael2021observing}. However, for the geometrically thick
disks as discussed in this work, this region may be obscured by
radiation from outside the equatorial plane, making it difficult to
distinguish. Compared with thin disks, thick disks are more
physically realistic, which indicates that direct imaging of the
black hole event horizon remains challenging.

In Fig.~\textbf{\ref{fig2}}, when $\theta_{o} \to 0^\circ$ (the
first column), the higher-order image appears as a perfect circular
ring. When $\theta_{o}$ increases to $17^\circ$ (the second column),
the higher-order image shifts downward on the screen. When
$\theta_{o}$ further increases to $75^\circ$ (the third column), the
intensity on the left and right sides of the higher-order image
becomes significantly greater than that in the vertical direction.
Meanwhile, $\hat{\gamma}$ has almost no effect on the shape of the
higher-order image, but increasing $\hat{\gamma}$ reduces its size.
Interestingly, for $\theta_{o}=75^\circ$ (the third column), two
dark regions appear inside the higher-order image, with the upper
one slightly dimmer than the lower one, a phenomenon arising from
gravitational lensing. In Fig. \textbf{\ref{fig3}}, we fixed
$\hat{\gamma}=0.95$ and study the effects of $\hat{\lambda}$ and
$\theta_o$ on the shadow images, which shows that the similar
results as discussed in Fig.~\textbf{\ref{fig2}}. The increasing
values of $\hat{\lambda}$ reduces the size of the higher-order
image, while increasing $\theta_o$ alters its shape and causes the
horizons outline to be obscured.

\begin{figure}[H]
\centering
\subfigure[$85\,\mathrm{GHz}$]{\includegraphics[scale=0.4]{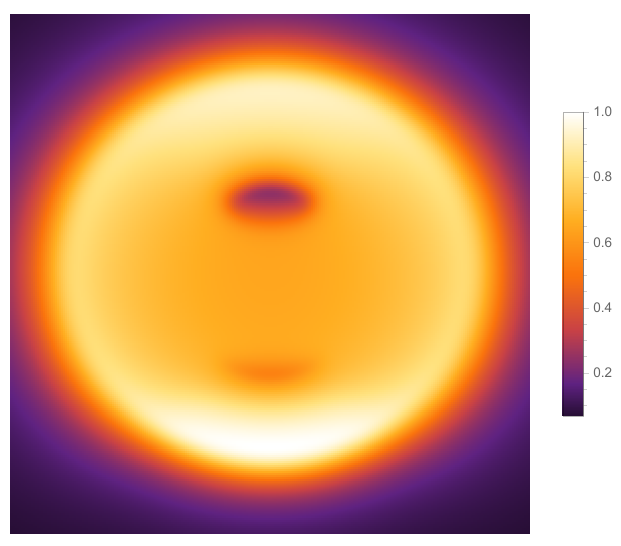}}
\subfigure[$230\,\mathrm{GHz}$]{\includegraphics[scale=0.4]{Iso9.pdf}}
\subfigure[$345\,\mathrm{GHz}$]{\includegraphics[scale=0.4]{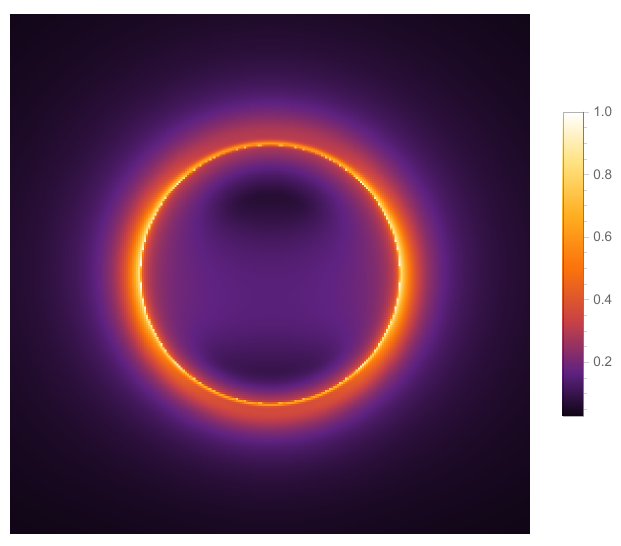}}
\caption{Effects of different observing frequencies on the
phenomenological model with isotropic radiation. The accretion flow
motion mode is the infalling motion, with fixed parameters
$\hat{\gamma}=0.95$, $\hat{\lambda}=0.9$, and
$\theta_{o}=75^\circ$.} \label{fig9}
\end{figure}

\begin{figure}[H]\centering
\subfigure[Orbiting motion]{\includegraphics[scale=0.4]{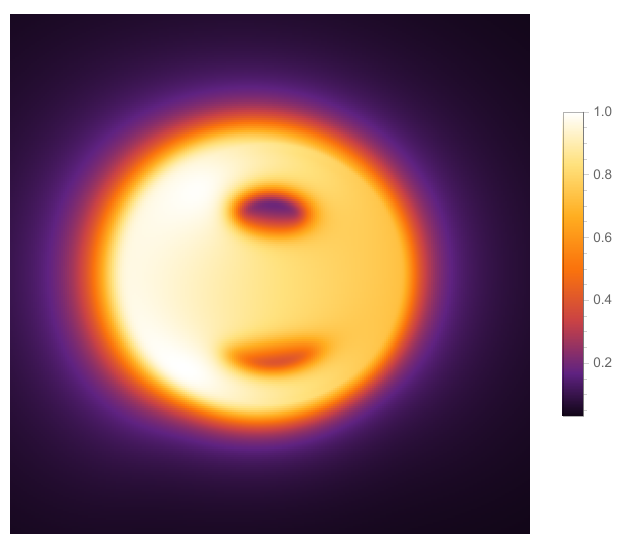}}
\subfigure[Infalling motion]{\includegraphics[scale=0.4]{Iso9.pdf}}
\subfigure[Combined motion]{\includegraphics[scale=0.4]{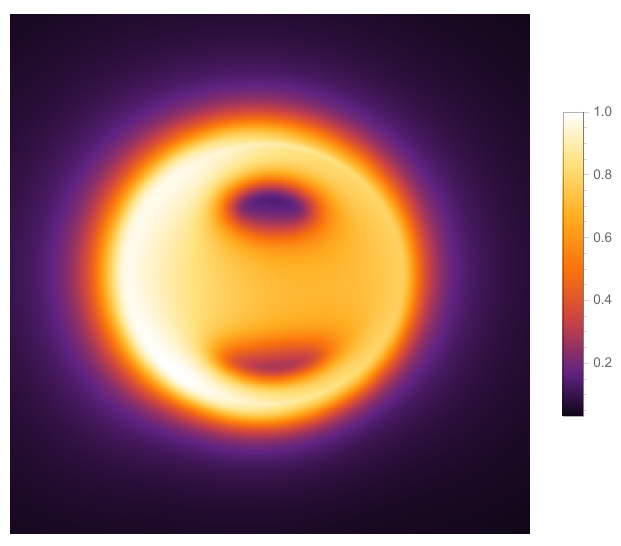}}
\caption{Effects of different accretion flow motion modes on the
phenomenological model with isotropic radiation. The parameters are
fixed at the observing frequency $230\,\mathrm{GHz}$,
$\hat{\gamma}=0.95$, $\hat{\lambda}=0.9$, and
$\theta_{o}=75^\circ$.}\label{fig10}
\end{figure}

In the previous analysis, the observing frequency was fixed at
$230\,\mathrm{GHz}$ and the accretion flow motion mode was chosen as
the infalling motion. Below we briefly analyze the effects of
different observing frequencies and accretion flow motion modes on
the phenomenological model with isotropic radiation. Figure
\textbf{\ref{fig9}} from left to right corresponds to observing
frequencies of $85\,\mathrm{GHz}$, $230\,\mathrm{GHz}$, and
$345\,\mathrm{GHz}$. As shown in the figure, when the observing
frequency is $85\,\mathrm{GHz}$, the outline of the event horizon is
completely obscured, and the primary and higher-order images are
difficult to distinguish. As the frequency increases, the intensity
decreases, and the higher-order image and the horizons outline
become clearly visible. Figure \textbf{\ref{fig10}} from left to
right corresponds to the accretion flow motion modes of the orbiting
motion, the infalling motion, and the combined motion, respectively
with $\beta_1=\beta_2=0.2$. The images of the orbiting motion and
the combined motion are similar, because $\beta_1=\beta_2=0.2$ makes
the contribution of the orbiting motion dominant in the combined
motion. This analysis shows that the orbiting motion causes the
horizons outline to be obscured by radiation from outside the
equatorial plane and significantly increases the intensity, while
the infalling motion makes the primary and higher-order images
easier to distinguish.

\subsection{Phenomenological Model with Anisotropic Radiation}

\begin{figure}[H]\centering
\subfigure[$\hat{\gamma}=0.001,\theta_{o}=0.001^\circ$]{\includegraphics[scale=0.4]{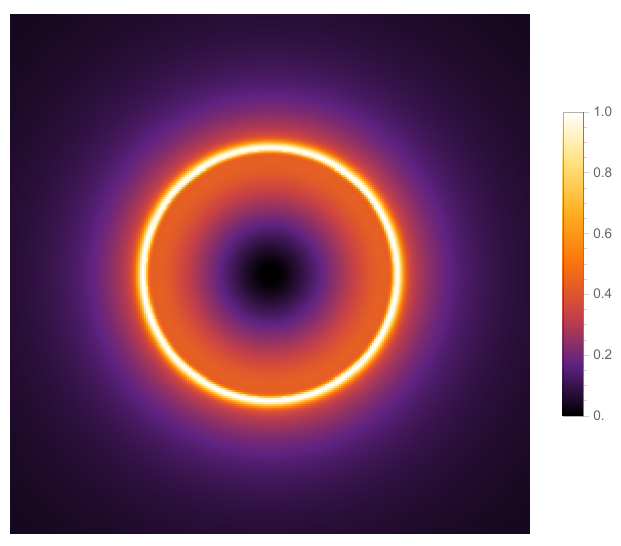}}
\subfigure[$\hat{\gamma}=0.001,\theta_{o}=17^\circ$]{\includegraphics[scale=0.4]{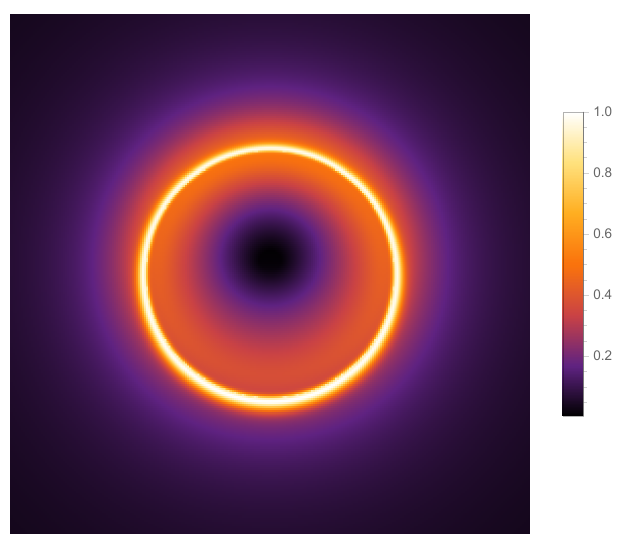}}
\subfigure[$\hat{\gamma}=0.001,\theta_{o}=75^\circ$]{\includegraphics[scale=0.4]{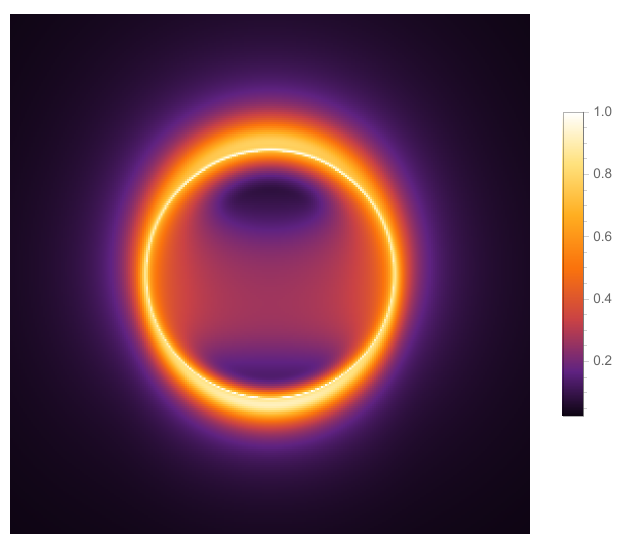}}
\subfigure[$\hat{\gamma}=0.55,\theta_{o}=0.001^\circ$]{\includegraphics[scale=0.4]{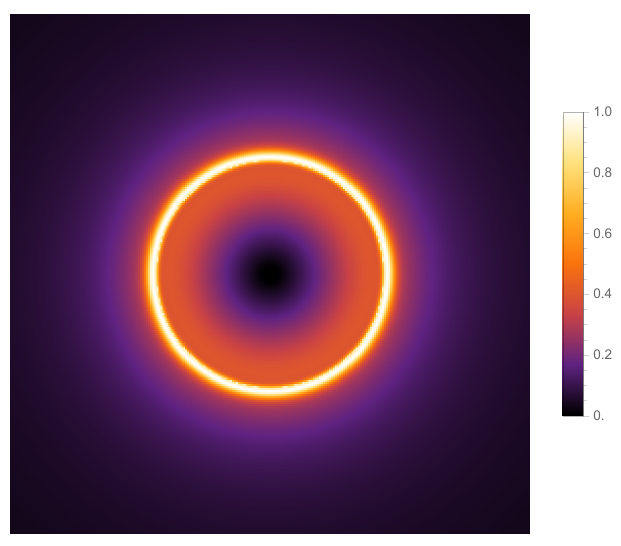}}
\subfigure[$\hat{\gamma}=0.55,\theta_{o}=17^\circ$]{\includegraphics[scale=0.4]{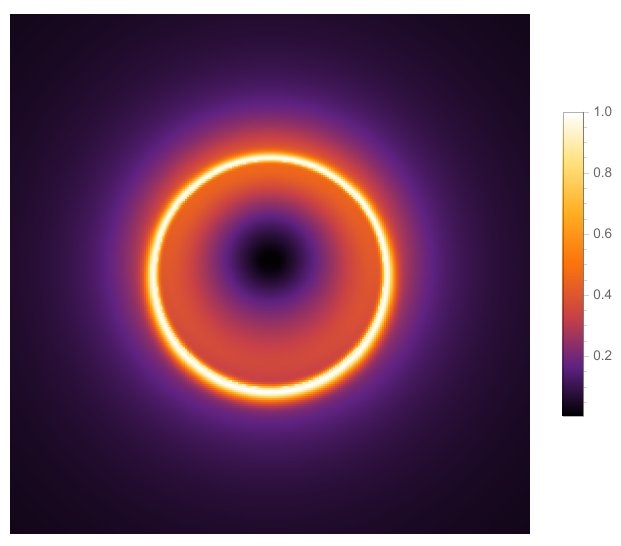}}
\subfigure[$\hat{\gamma}=0.55,\theta_{o}=75^\circ$]{\includegraphics[scale=0.4]{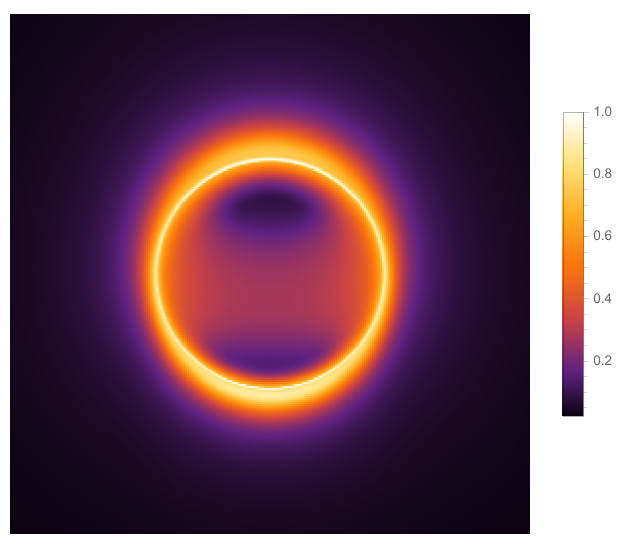}}
\subfigure[$\hat{\gamma}=0.95,\theta_{o}=0.001^\circ$]{\includegraphics[scale=0.4]{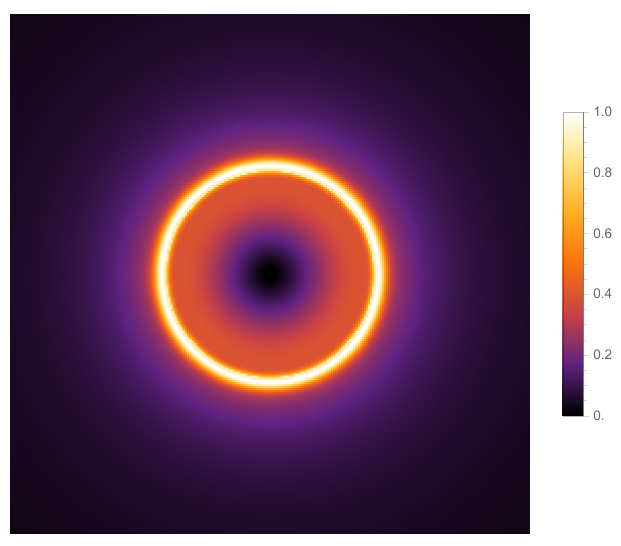}}
\subfigure[$\hat{\gamma}=0.95,\theta_{o}=17^\circ$]{\includegraphics[scale=0.4]{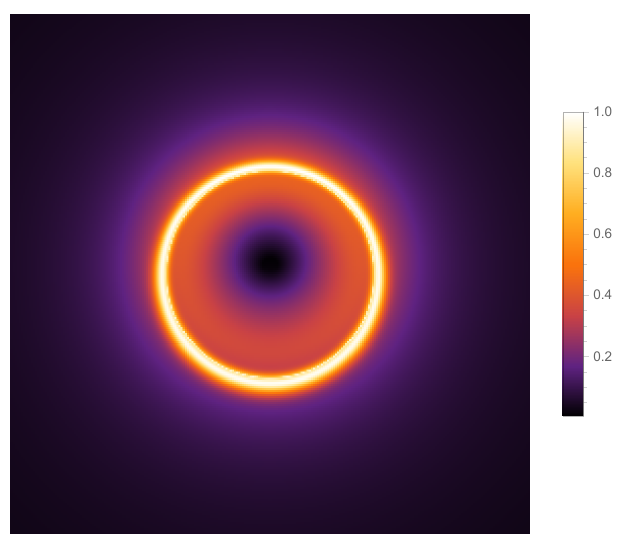}}
\subfigure[$\hat{\gamma}=0.95,\theta_{o}=75^\circ$]{\includegraphics[scale=0.4]{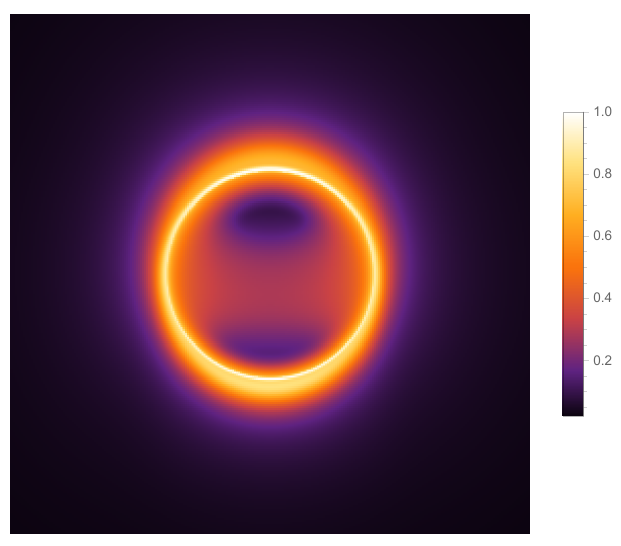}}
\caption{Black hole shadow images of the phenomenological model with
anisotropic radiation. The accretion flow motion mode is the
infalling motion. The parameters are fixed at $\hat{\lambda}=0.9$
and the observing frequency is $230\,\mathrm{GHz}$.} \label{fig4}
\end{figure}

\begin{figure}[H]
\centering
\subfigure[$\hat{\lambda}=0.1,\theta_{o}=0.001^\circ$]{\includegraphics[scale=0.4]{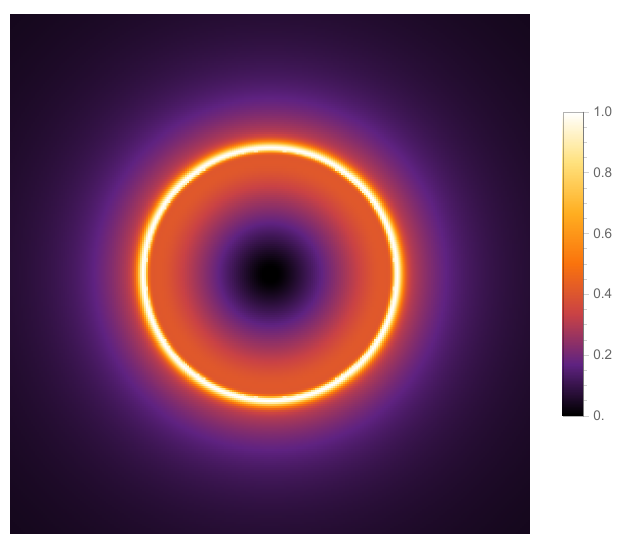}}
\subfigure[$\hat{\lambda}=0.1,\theta_{o}=17^\circ$]{\includegraphics[scale=0.4]{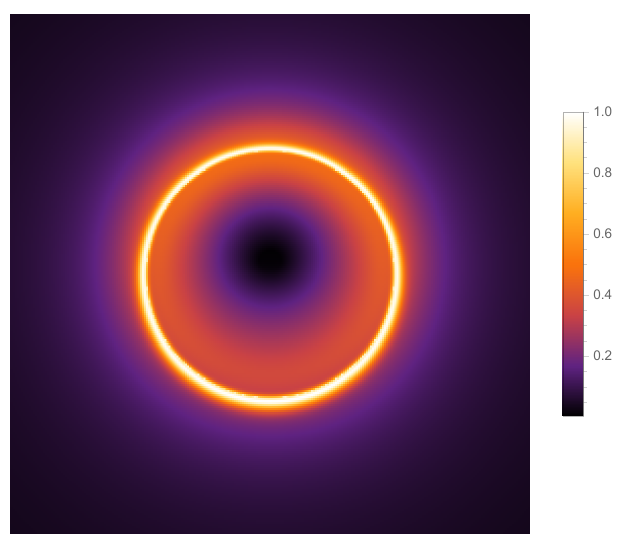}}
\subfigure[$\hat{\lambda}=0.1,\theta_{o}=75^\circ$]{\includegraphics[scale=0.4]{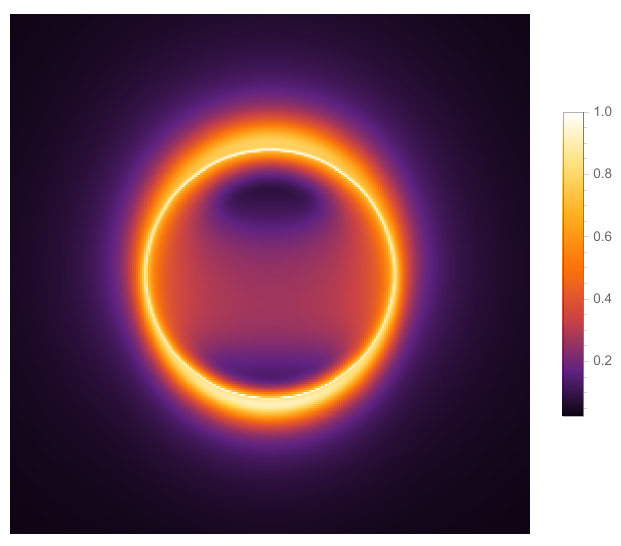}}
\subfigure[$\hat{\lambda}=0.55,\theta_{o}=0.001^\circ$]{\includegraphics[scale=0.4]{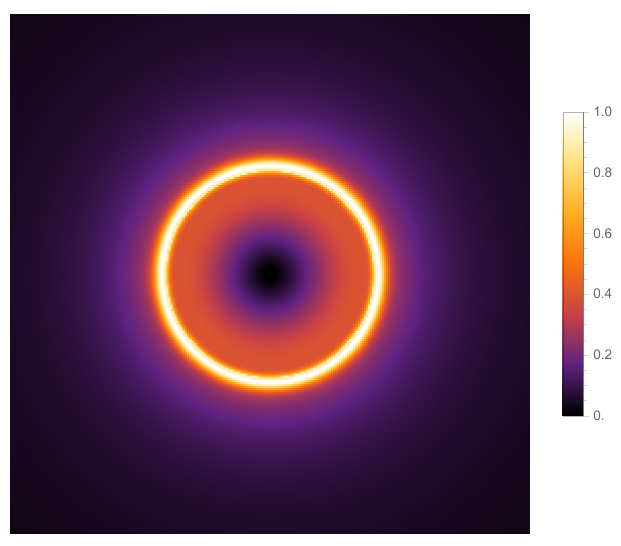}}
\subfigure[$\hat{\lambda}=0.55,\theta_{o}=17^\circ$]{\includegraphics[scale=0.4]{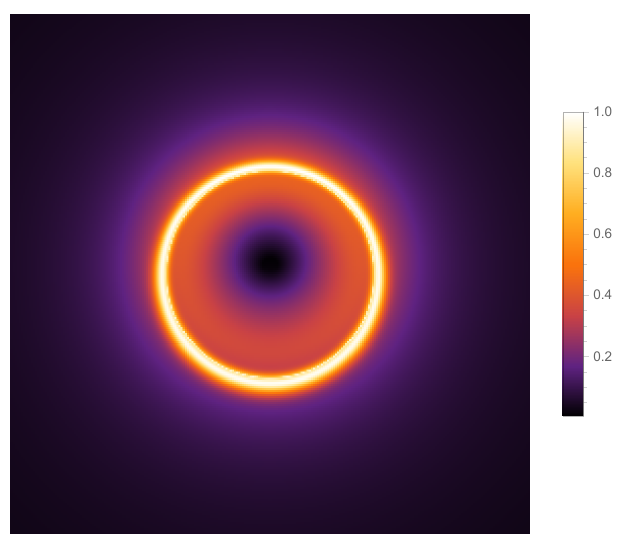}}
\subfigure[$\hat{\lambda}=0.55,\theta_{o}=75^\circ$]{\includegraphics[scale=0.4]{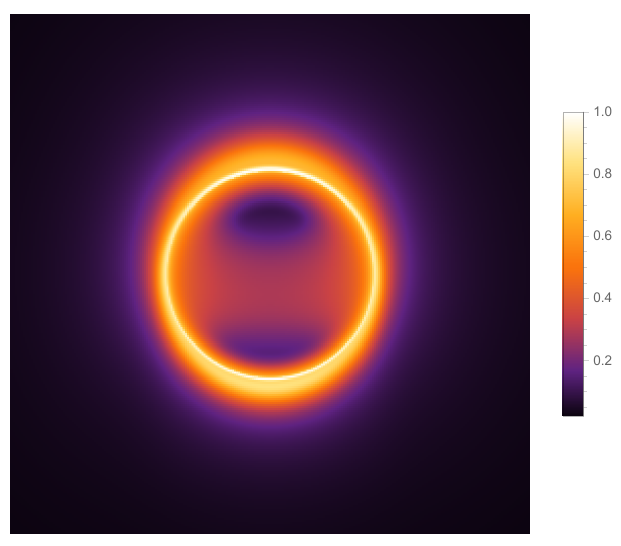}}
\subfigure[$\hat{\lambda}=0.9,\theta_{o}=0.001^\circ$]{\includegraphics[scale=0.4]{Ani7.pdf}}
\subfigure[$\hat{\lambda}=0.9,\theta_{o}=17^\circ$]{\includegraphics[scale=0.4]{Ani8.pdf}}
\subfigure[$\hat{\lambda}=0.9,\theta_{o}=75^\circ$]{\includegraphics[scale=0.4]{Ani9.pdf}}
\caption{Black hole shadow images of the phenomenological model with
anisotropic radiation. The accretion flow motion mode is the
infalling motion. The parameters are fixed at $\hat{\gamma}=0.95$
and the observing frequency is $230\,\mathrm{GHz}$.}\label{fig5}
\end{figure}

Now, we will discuss the significant properties of black hole shadow
images, which are produced by the phenomenological model with an
anisotropic radiation framework, as depicted in Figs.
\textbf{\ref{fig4}} and \textbf{\ref{fig5}}. The observing frequency
is $230 \,\mathrm{GHz}$, and the accretion flow motion mode is
chosen as the infalling motion. Figure \textbf{\ref{fig4}}
illustrates the effect of $\hat{\lambda}$, while
Fig.~\textbf{\ref{fig5}} illustrates the effect of $\hat{\gamma}$.
From these images it can be seen that, under anisotropic radiation,
the influences of $\hat{\lambda}$, $\hat{\gamma}$, and $\theta_{o}$
on the black hole shadow are similar to those under isotropic
radiation: increasing $\hat{\lambda}$ and $\hat{\gamma}$ reduces the
size of the higher-order image, and increasing $\theta_{o}$ changes
the shape of the higher-order image. The difference is that when
$\theta_{o}=75^\circ$, the higher-order image remains approximately
circular under isotropic radiation, but under anisotropic radiation
it is significantly deformed into an elliptical shape. This analysis
indicates that the closer the observer is to the equatorial plane,
the more significant the effect of the radiation type on the shadow
image.

\subsection{HOU Disk Model}
\begin{figure}[H]
\centering
\subfigure[$\hat{\gamma}=0.001,\theta_{o}=0.001^\circ$]{\includegraphics[scale=0.4]{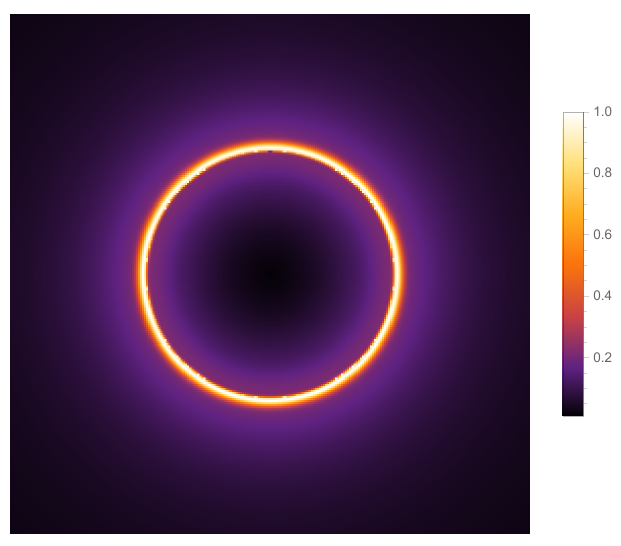}}
\subfigure[$\hat{\gamma}=0.001,\theta_{o}=17^\circ$]{\includegraphics[scale=0.4]{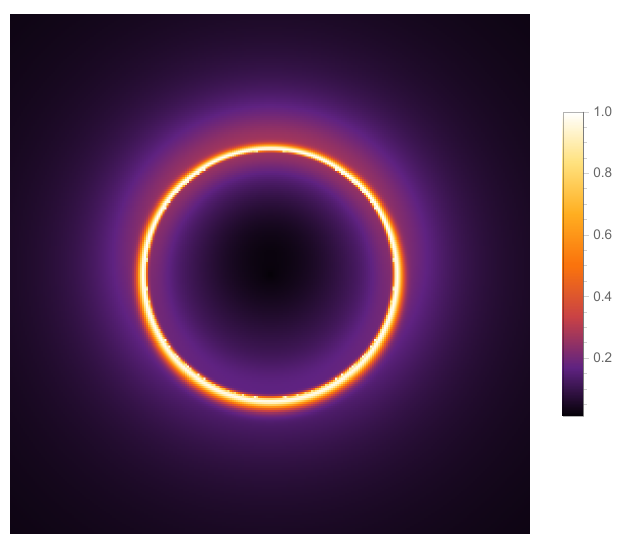}}
\subfigure[$\hat{\gamma}=0.001,\theta_{o}=75^\circ$]{\includegraphics[scale=0.4]{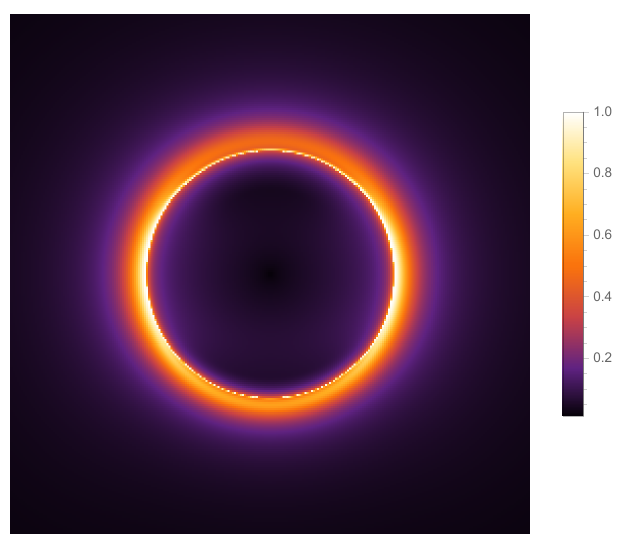}}
\subfigure[$\hat{\gamma}=0.55,\theta_{o}=0.001^\circ$]{\includegraphics[scale=0.4]{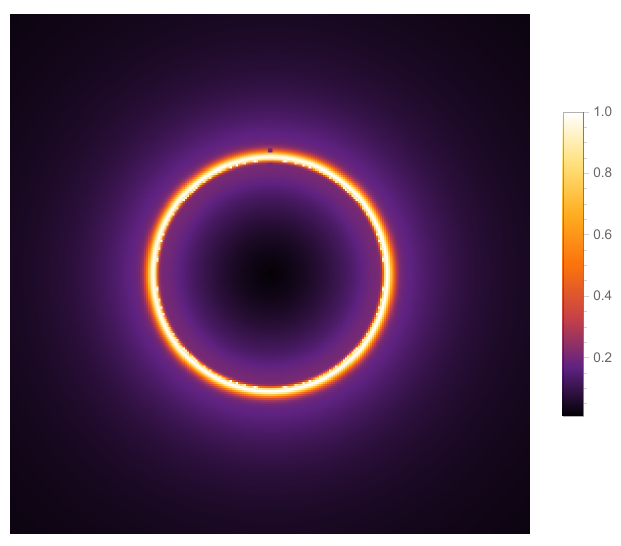}}
\subfigure[$\hat{\gamma}=0.55,\theta_{o}=17^\circ$]{\includegraphics[scale=0.4]{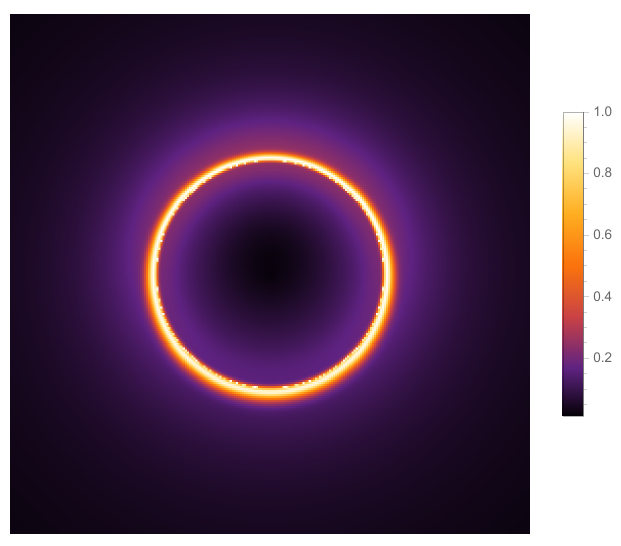}}
\subfigure[$\hat{\gamma}=0.55,\theta_{o}=75^\circ$]{\includegraphics[scale=0.4]{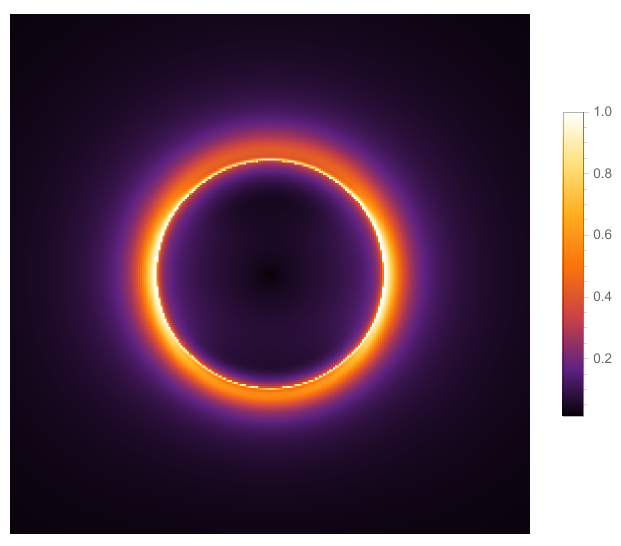}}
\subfigure[$\hat{\gamma}=0.95,\theta_{o}=0.001^\circ$]{\includegraphics[scale=0.4]{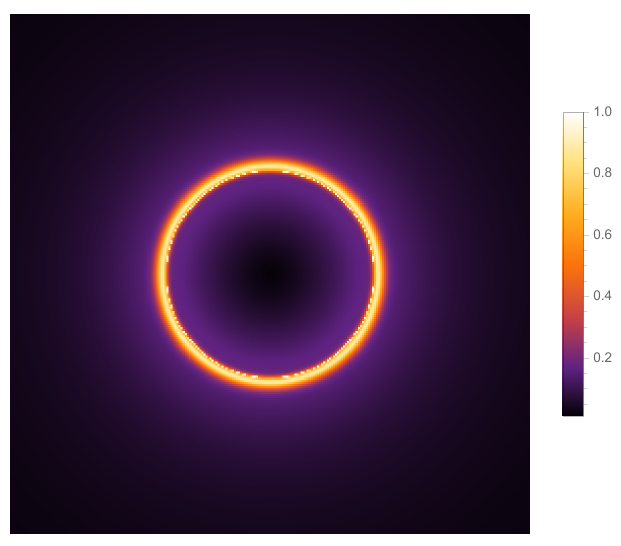}}
\subfigure[$\hat{\gamma}=0.95,\theta_{o}=17^\circ$]{\includegraphics[scale=0.4]{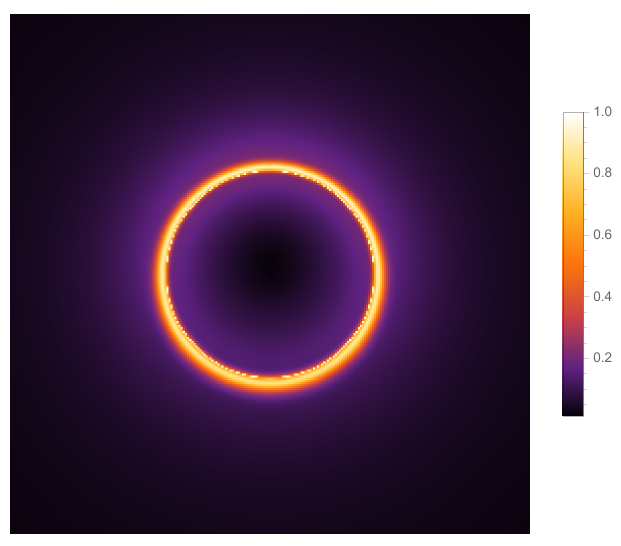}}
\subfigure[$\hat{\gamma}=0.95,\theta_{o}=75^\circ$]{\includegraphics[scale=0.4]{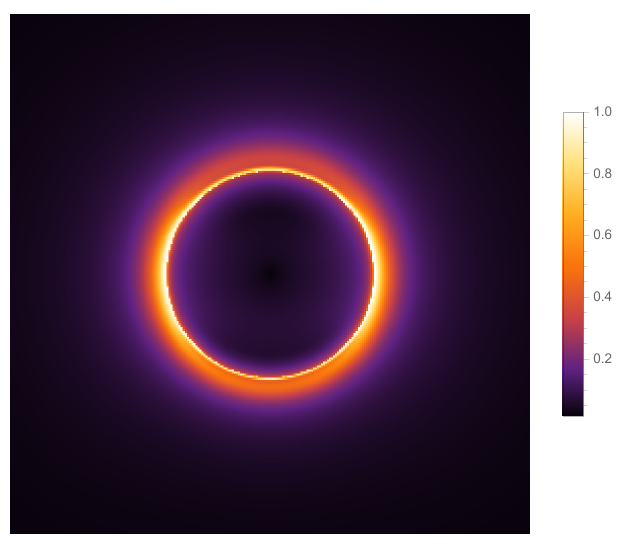}}
\caption{Black hole shadow images of the HOU disk model. The
accretion flow motion mode is the infalling motion. The parameters
are fixed at $\hat{\lambda}=0.9$ and the observing frequency is
$230\,\mathrm{GHz}$.} \label{fig6}
\end{figure}

\begin{figure}[H]
\centering
\subfigure[$\hat{\lambda}=0.1,\theta_{o}=0.001^\circ$]{\includegraphics[scale=0.4]{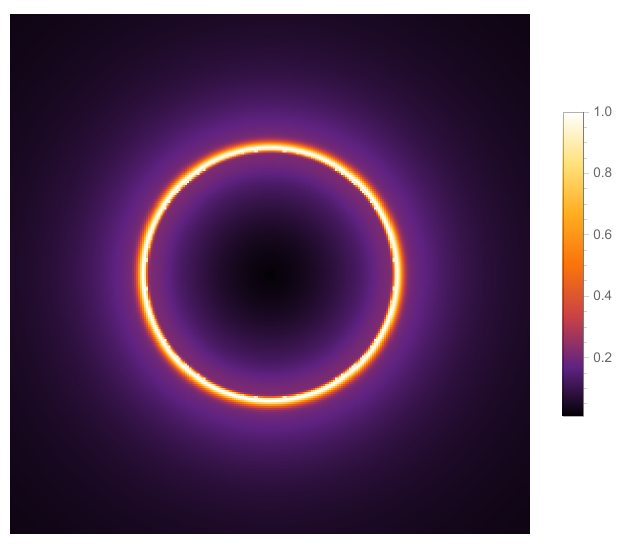}}
\subfigure[$\hat{\lambda}=0.1,\theta_{o}=17^\circ$]{\includegraphics[scale=0.4]{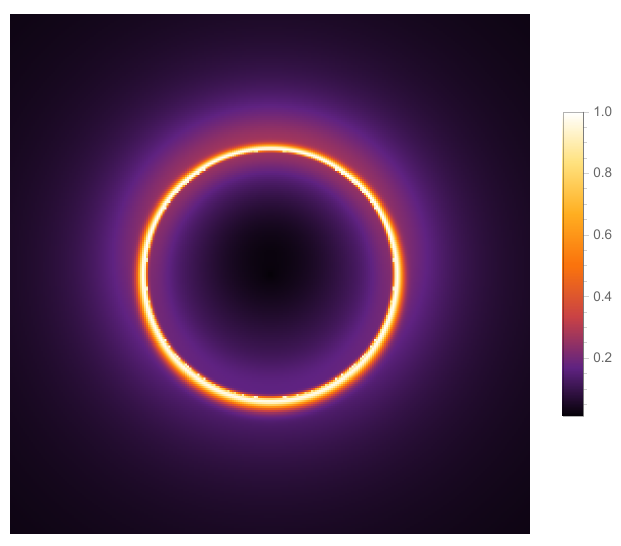}}
\subfigure[$\hat{\lambda}=0.1,\theta_{o}=75^\circ$]{\includegraphics[scale=0.4]{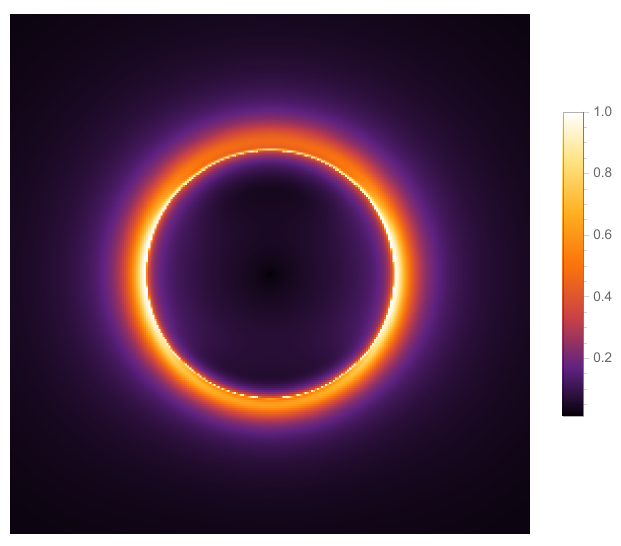}}
\subfigure[$\hat{\lambda}=0.55,\theta_{o}=0.001^\circ$]{\includegraphics[scale=0.4]{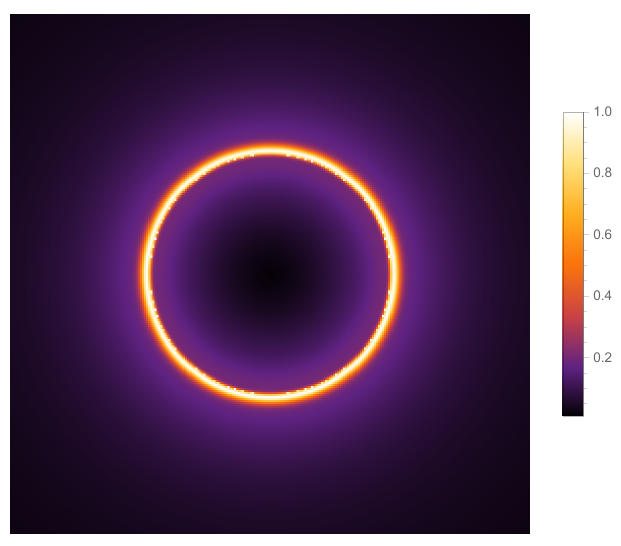}}
\subfigure[$\hat{\lambda}=0.55,\theta_{o}=17^\circ$]{\includegraphics[scale=0.4]{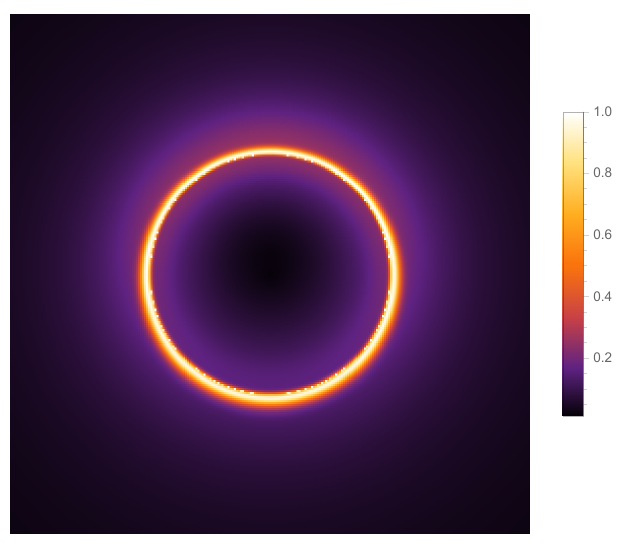}}
\subfigure[$\hat{\lambda}=0.55,\theta_{o}=75^\circ$]{\includegraphics[scale=0.4]{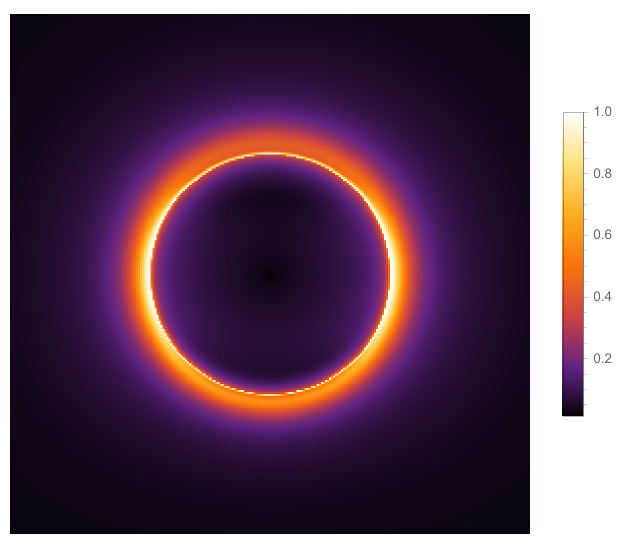}}
\subfigure[$\hat{\lambda}=0.9,\theta_{o}=0.001^\circ$]{\includegraphics[scale=0.4]{Hou7.pdf}}
\subfigure[$\hat{\lambda}=0.9,\theta_{o}=17^\circ$]{\includegraphics[scale=0.4]{Hou8.pdf}}
\subfigure[$\hat{\lambda}=0.9,\theta_{o}=75^\circ$]{\includegraphics[scale=0.4]{Hou9.pdf}}
\caption{Black hole shadow images of the HOU disk model. The
accretion flow motion mode is the infalling motion. The parameters
are fixed at $\hat{\gamma}=0.95$ and the observing frequency is
$230\,\mathrm{GHz}$.} \label{fig7}
\end{figure}

Considering the mechanism of the HOU disk model, in Figs
\textbf{\ref{fig6}} and \textbf{\ref{fig7}}, we shows that the black
hole shadow images with anisotropic radiation. The observing
frequency is $230 \,\mathrm{GHz}$, and the accretion flow motion
mode is chosen as the infalling motion. Figure \textbf{\ref{fig6}}
illustrates the effect of $\hat{\lambda}$, while
Fig.~\textbf{\ref{fig7}} illustrates the effect of $\hat{\gamma}$.
The bright ring in the images represents the higher-order image, and
the dark region inside the higher-order image originates from the
event horizon. Increasing $\hat{\lambda}$ or $\hat{\gamma}$ reduces
the size of the higher-order image. Increasing $\theta_{o}$ enhances
the brightness of the primary image outside the higher-order image
but hardly changes the size of the higher-order image, which is in
sharp contrast to the phenomenological model. Meanwhile, for the HOU
disk model, at high observer inclinations, the obscuration of the
horizons outline by radiation from outside the equatorial plane is
weakened.

\begin{figure}[H]\centering
\subfigure[Phenomenological model, isotropic
radiation]{\includegraphics[scale=0.4]{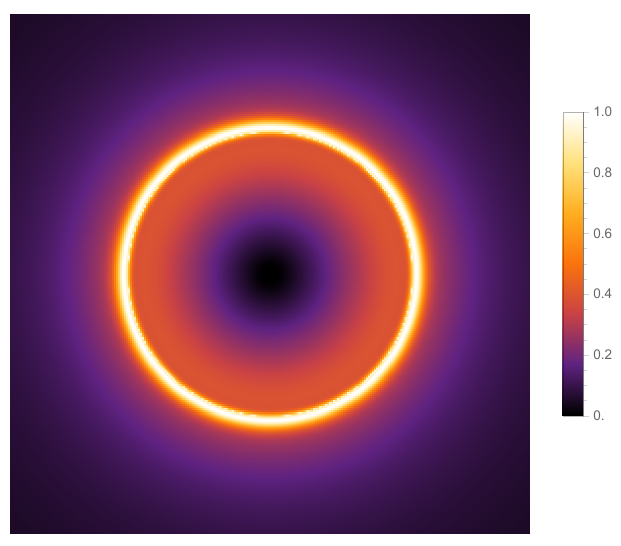}}
\subfigure[Phenomenological model, anisotropic
radiation]{\includegraphics[scale=0.4]{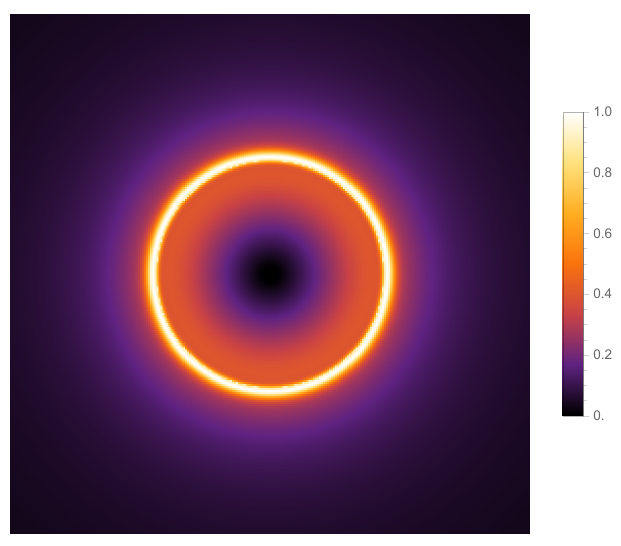}} \subfigure[HOU
disk model]{\includegraphics[scale=0.4]{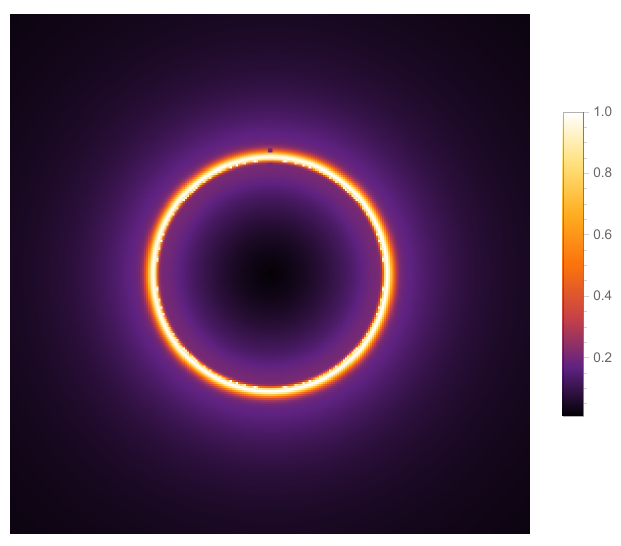}}
\subfigure[Phenomenological model, isotropic
radiation]{\includegraphics[scale=0.4]{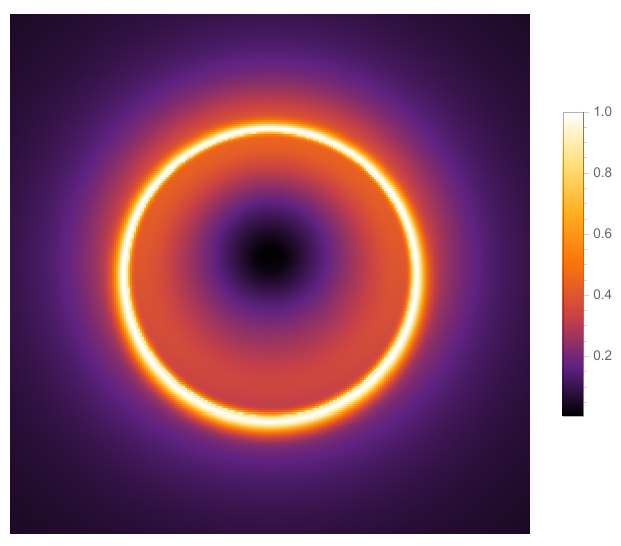}}
\subfigure[Phenomenological model, anisotropic
radiation]{\includegraphics[scale=0.4]{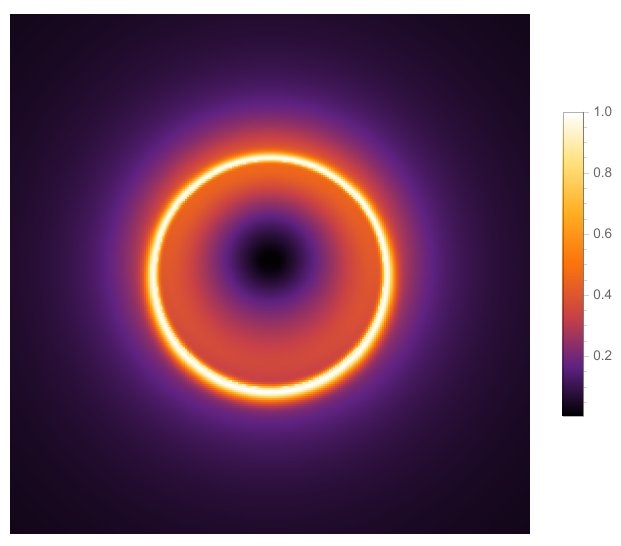}} \subfigure[HOU
disk model]{\includegraphics[scale=0.4]{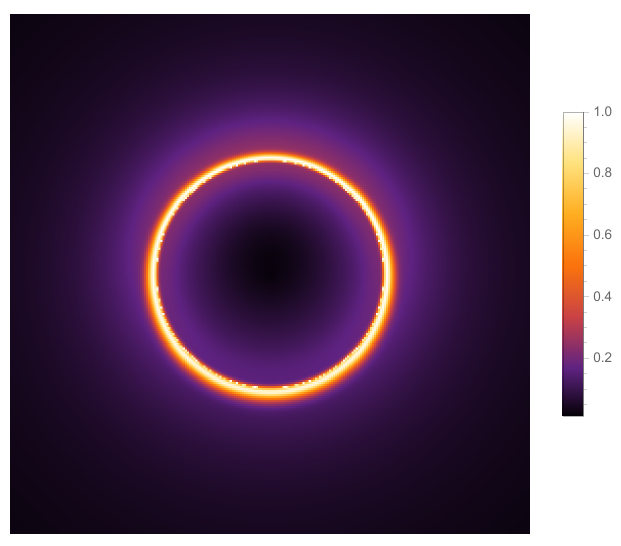}}
\subfigure[Phenomenological model, isotropic
radiation]{\includegraphics[scale=0.4]{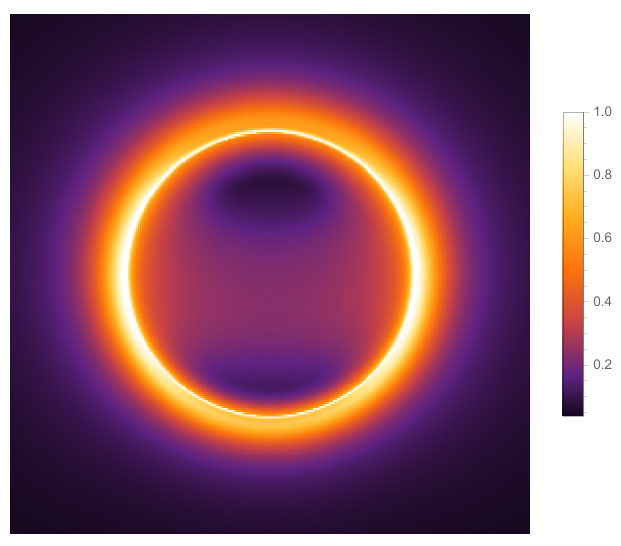}}
\subfigure[Phenomenological model, anisotropic
radiation]{\includegraphics[scale=0.4]{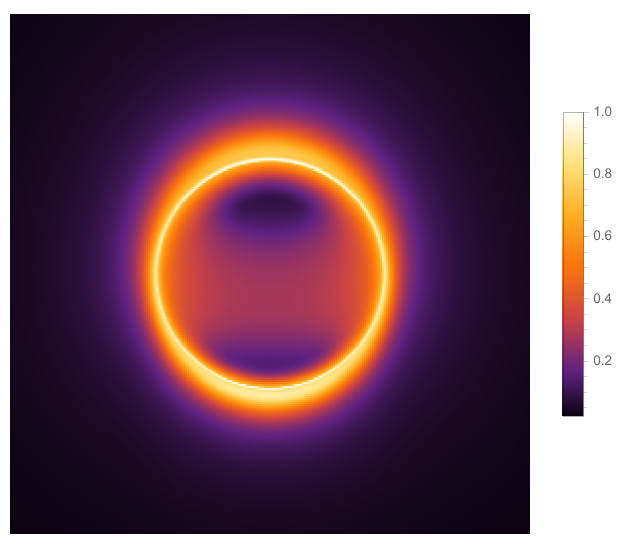}} \subfigure[HOU
disk model]{\includegraphics[scale=0.4]{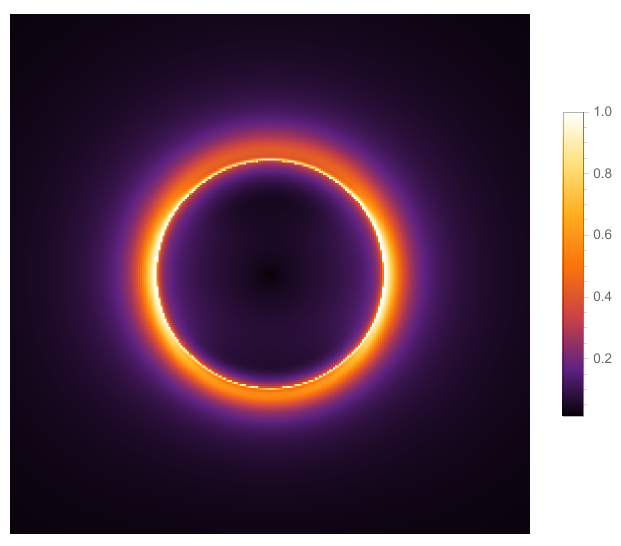}} \caption{Effects
of different observer inclinations on the black hole shadow for
various accretion disk models. The first to third rows correspond to
observer inclinations $\theta_{o}=0.001^\circ$, $17^\circ$, and
$75^\circ$, respectively. The accretion flow motion mode is the
infalling motion, with fixed parameters observing frequency
$230\,\mathrm{GHz}$, $\hat{\gamma}=0.55$, and
$\hat{\lambda}=0.9$.}\label{fig8}
\end{figure}

To directly compare the differences among the three models discussed
above, Fig.~\textbf{\ref{fig8}} shows the effects of different
accretion disk models on the black hole shadow at various observer
inclinations. It can be seen that, for the same observer
inclination, the phenomenological model with anisotropic radiation
corresponds to the largest size of the higher-order image. The
brightness of the primary image in the phenomenological model is
significantly greater than that in the HOU disk model. Meanwhile, at
higher observer inclinations, such as when $\theta_{o}=75^\circ$,
(see third row), the obscuration of the horizons outline by
radiation from outside the equatorial plane is more pronounced in
the phenomenological model, indicating that the gravitational
lensing effect is stronger in this case.

\subsection{Polarization Images in the HOU Disk Model}

\begin{figure}[H]
\centering \subfigure[Stokes parameter
$\mathcal{I}_{o}$]{\includegraphics[scale=0.6]{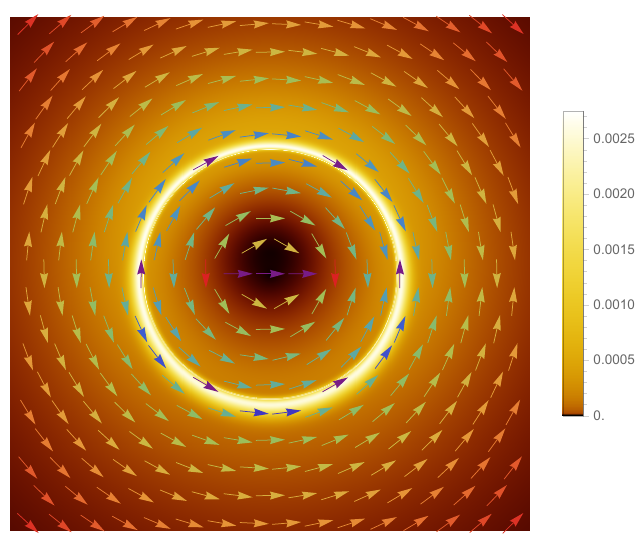}}
\subfigure[Stokes parameter
$\mathcal{Q}_{o}$]{\includegraphics[scale=0.6]{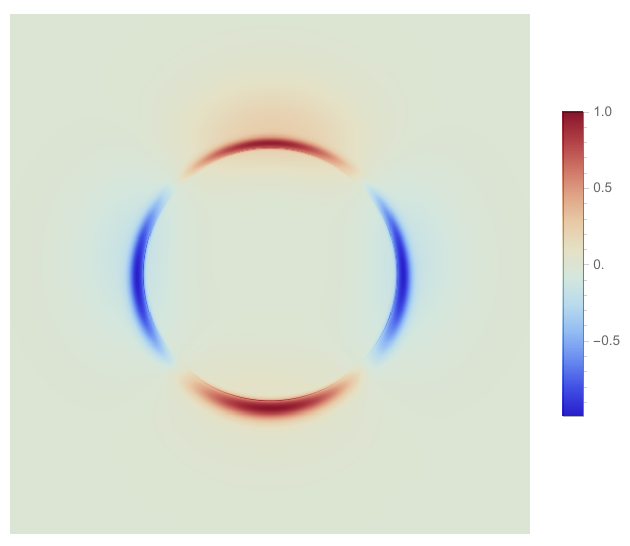}}
\subfigure[Stokes parameter
$\mathcal{U}_{o}$]{\includegraphics[scale=0.6]{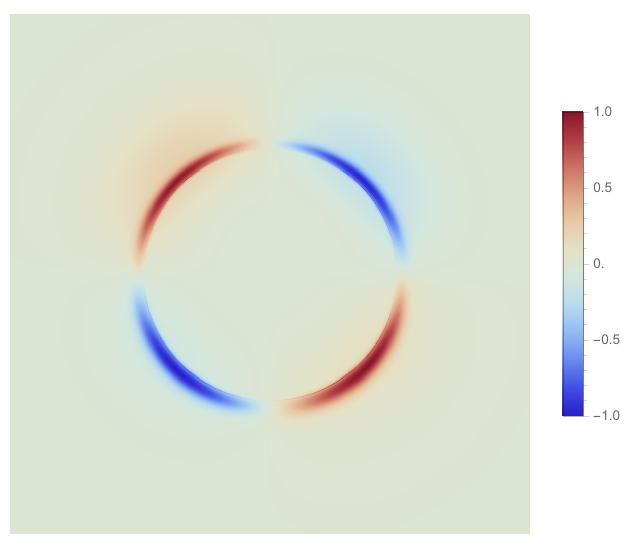}}
\subfigure[Stokes parameter
$\mathcal{V}_{o}$]{\includegraphics[scale=0.6]{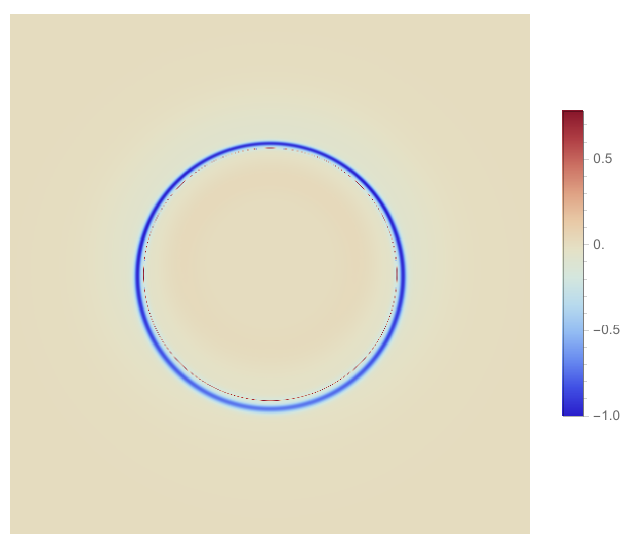}}
\caption{Stokes parameters $\mathcal{I}_{o}, \mathcal{Q}_{o},
\mathcal{U}_{o}, \mathcal{V}_{o}$ in the HOU disk model. The
accretion flow follows the infalling motion, with fixed parameters
$\hat{\lambda}=0.9$, $\hat{\gamma}=0.55$, $\theta_o=17^\circ$, and
observing frequency is $230\,\mathrm{GHz}$.} \label{fig11}
\end{figure}

In Fig.~\textbf{\ref{fig11}}, we presents the numerical results of
the Stokes parameters $\mathcal{I}_{o}, \mathcal{Q}_{o},
\mathcal{U}_{o}, \mathcal{V}_{o}$. In this case, the accretion flow
motion mode is the infalling motion, with fixed parameters
$\hat{\lambda}=0.9$, $\hat{\gamma}=0.55$, $\theta_o=17^\circ$, and
the observing frequency $230\,\mathrm{GHz}$. Here $\mathcal{I}_{o}$
reflects the intensity distribution, and the arrows denote the
polarization vector $\vec{f}$, which is calculated through
Eq.~(\ref{eq:pv}). The length and color depth of the arrows indicate
the polarization intensity $P_{o}$, while their direction
corresponds to the electric vector position angle
$\Phi_{\mathrm{EVPA}}$. Since $\vec{f}$ is always perpendicular to
the magnetic field $\vec{B}$, it can be inferred that the magnetic
field is approximately radial. By combining $\mathcal{Q}_{o}$ and
$\mathcal{U}_{o}$, the orientation of the electric vector $\vec{E}$
can be qualitatively determined, while $\mathcal{V}_{o}<0$ indicates
right-handed polarization. The parameters $\mathcal{Q}_{o}$,
$\mathcal{U}_{o}$, and $\mathcal{V}_{o}$ reach their peaks in the
region of the higher-order image and rapidly decay to zero away from
this region. Considering that $\mathcal{I}_{o}$ already contains the
complete information of the polarization image, only
$\mathcal{I}_{o}$ will be analyzed in the following.

\begin{figure}[H]
\centering
\subfigure[$\hat{\gamma}=0.001,\theta_{o}=0.001^\circ$]{\includegraphics[scale=0.4]{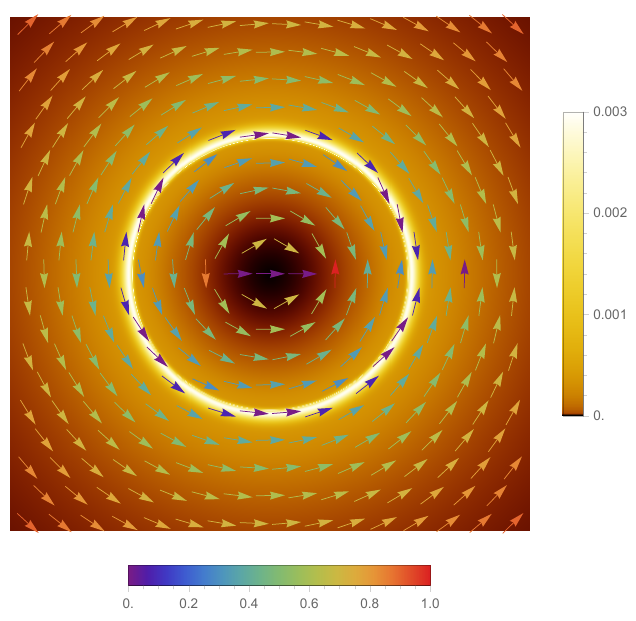}}
\subfigure[$\hat{\gamma}=0.001,\theta_{o}=17^\circ$]{\includegraphics[scale=0.4]{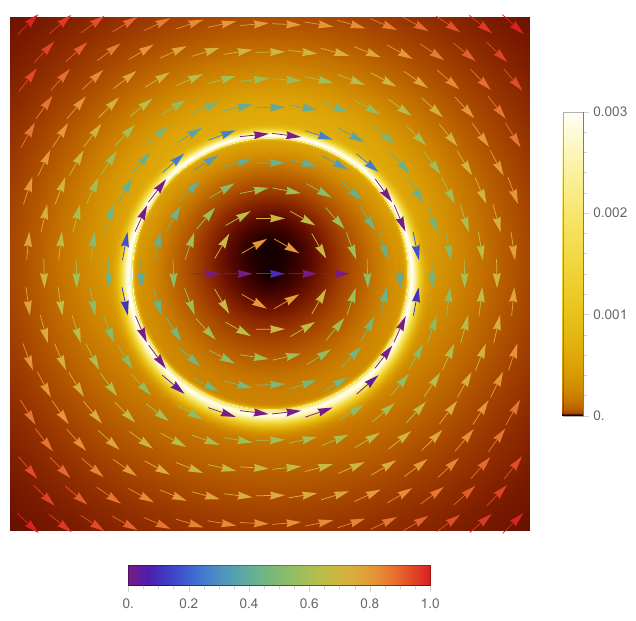}}
\subfigure[$\hat{\gamma}=0.001,\theta_{o}=75^\circ$]{\includegraphics[scale=0.4]{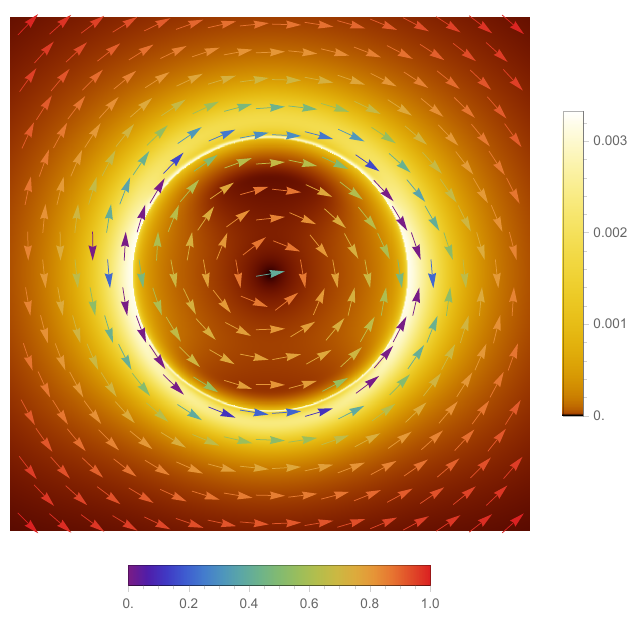}}
\subfigure[$\hat{\gamma}=0.55,\theta_{o}=0.001^\circ$]{\includegraphics[scale=0.4]{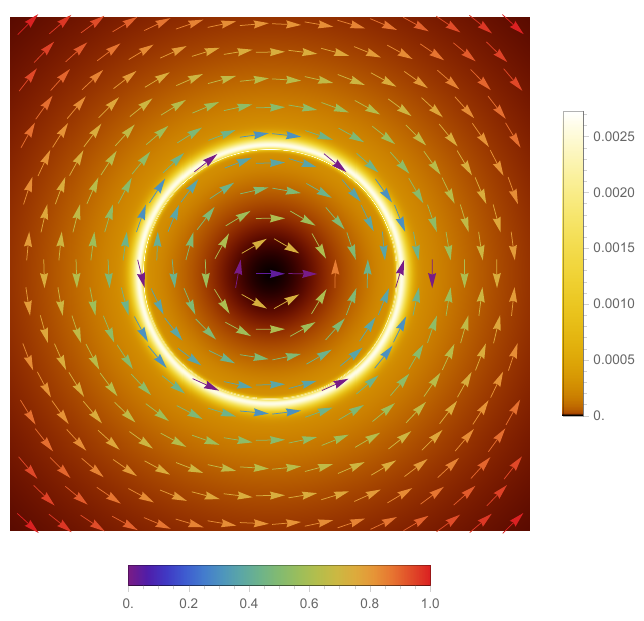}}
\subfigure[$\hat{\gamma}=0.55,\theta_{o}=17^\circ$]{\includegraphics[scale=0.4]{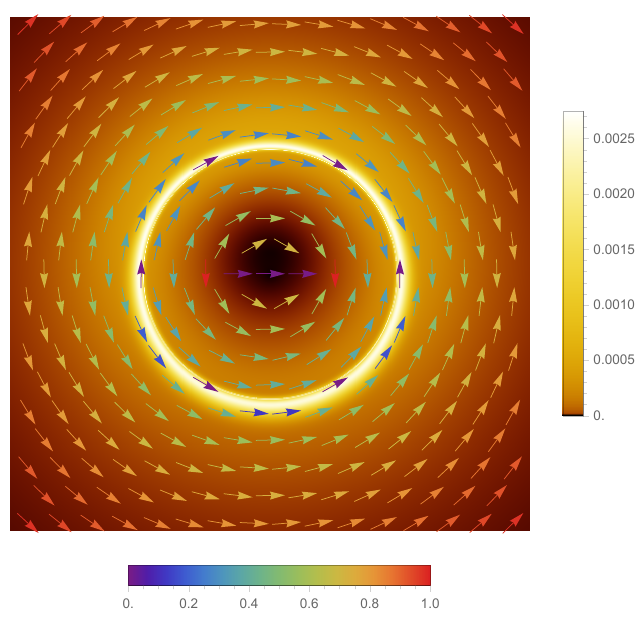}}
\subfigure[$\hat{\gamma}=0.55,\theta_{o}=75^\circ$]{\includegraphics[scale=0.4]{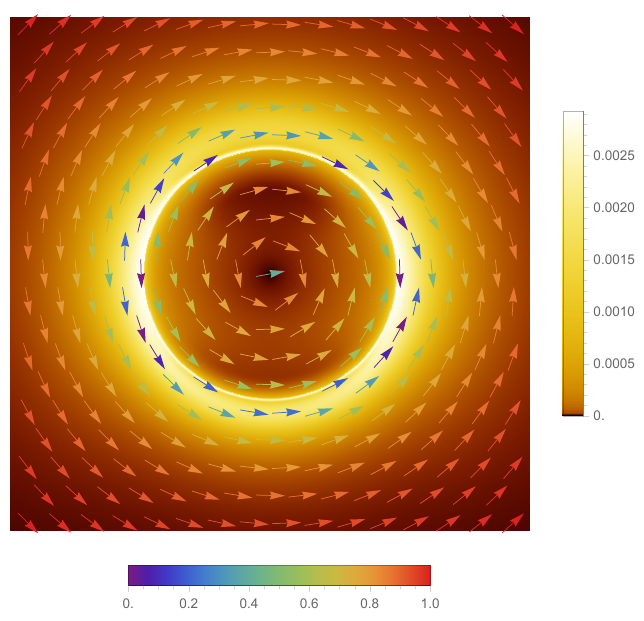}}
\subfigure[$\hat{\gamma}=0.95,\theta_{o}=0.001^\circ$]{\includegraphics[scale=0.4]{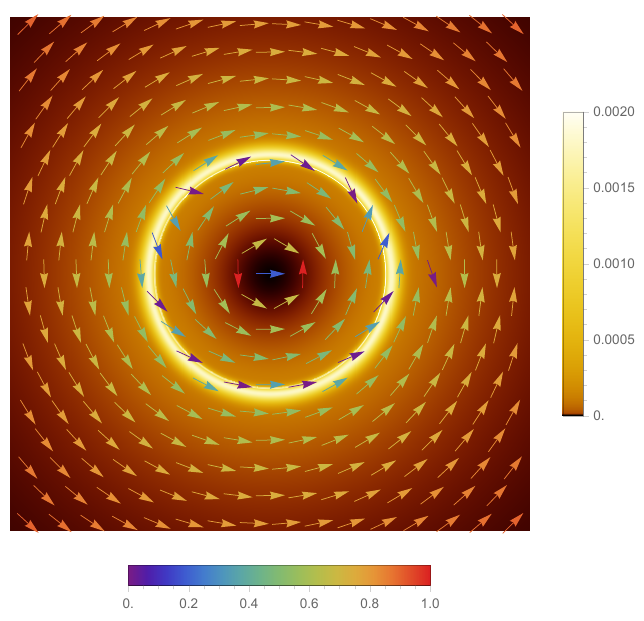}}
\subfigure[$\hat{\gamma}=0.95,\theta_{o}=17^\circ$]{\includegraphics[scale=0.4]{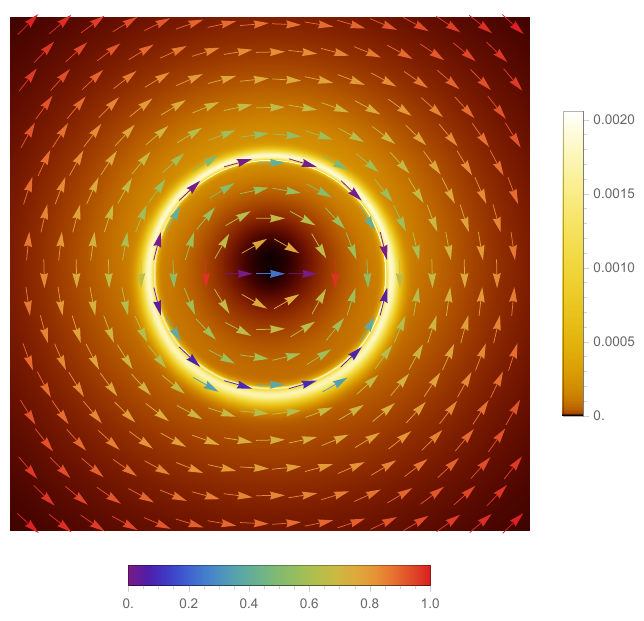}}
\subfigure[$\hat{\gamma}=0.95,\theta_{o}=75^\circ$]{\includegraphics[scale=0.4]{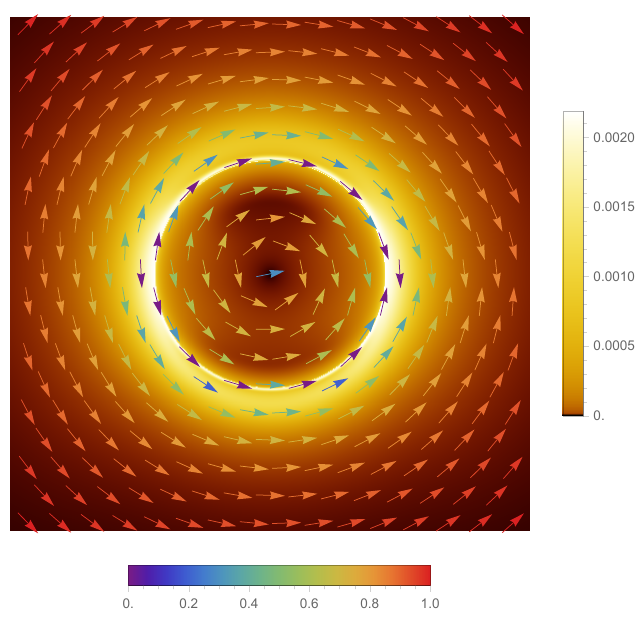}}
\caption{Polarized images of black hole shadow in the HOU disk
model. The accretion flow follows the infalling motion, with fixed
parameters $\hat{\lambda}=0.9$ and observing frequency is
$230\,\mathrm{GHz}$.} \label{fig12}
\end{figure}

\begin{figure}[H]
\centering
\subfigure[$\hat{\lambda}=0.1,\theta_{o}=0.001^\circ$]{\includegraphics[scale=0.4]{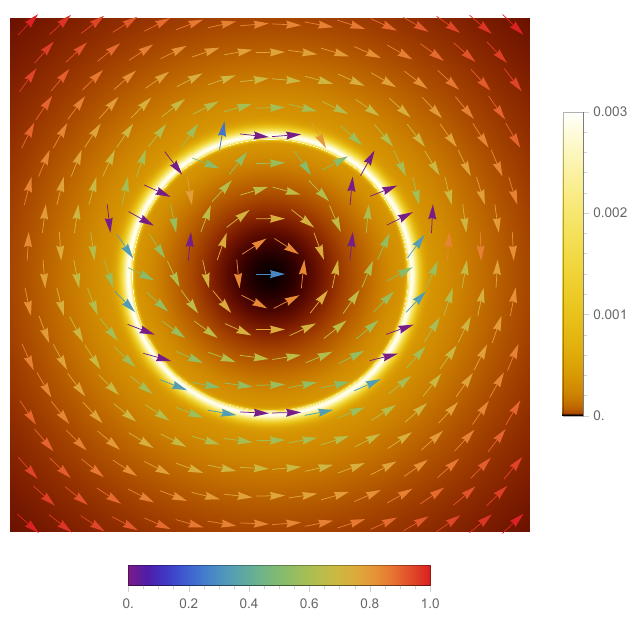}}
\subfigure[$\hat{\lambda}=0.1,\theta_{o}=17^\circ$]{\includegraphics[scale=0.4]{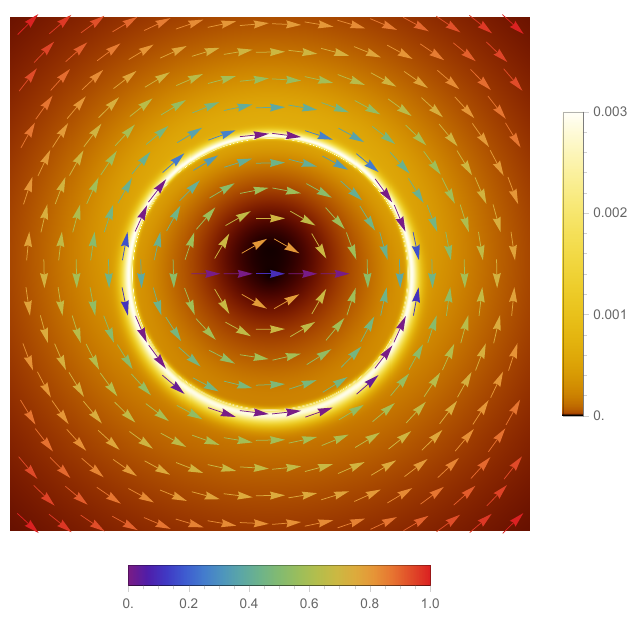}}
\subfigure[$\hat{\lambda}=0.1,\theta_{o}=75^\circ$]{\includegraphics[scale=0.4]{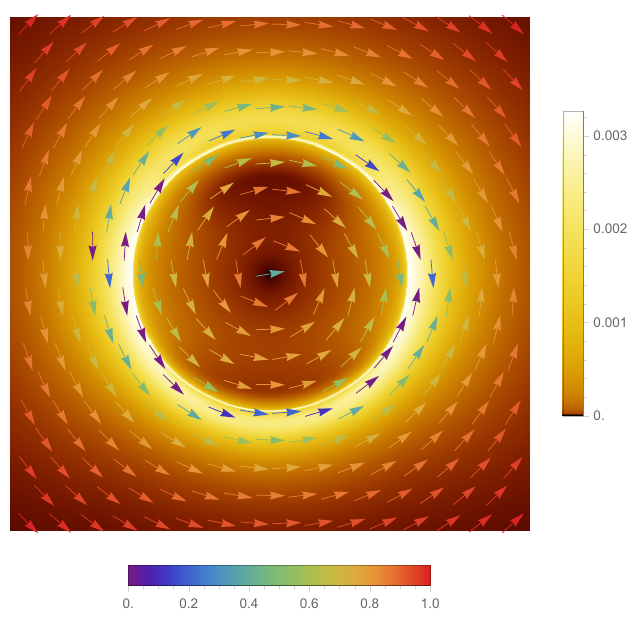}}
\subfigure[$\hat{\lambda}=0.55,\theta_{o}=0.001^\circ$]{\includegraphics[scale=0.4]{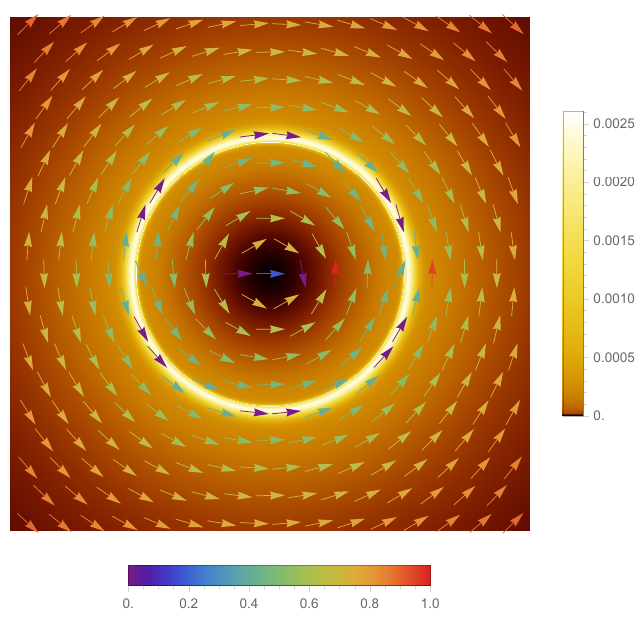}}
\subfigure[$\hat{\lambda}=0.55,\theta_{o}=17^\circ$]{\includegraphics[scale=0.4]{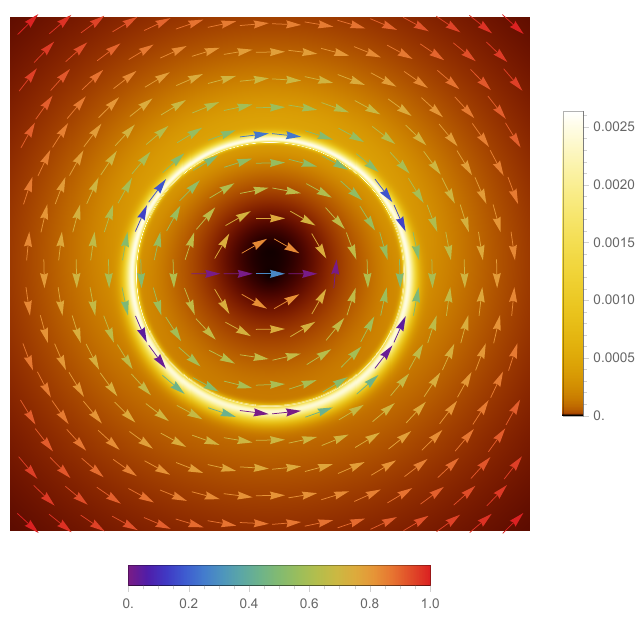}}
\subfigure[$\hat{\lambda}=0.55,\theta_{o}=75^\circ$]{\includegraphics[scale=0.4]{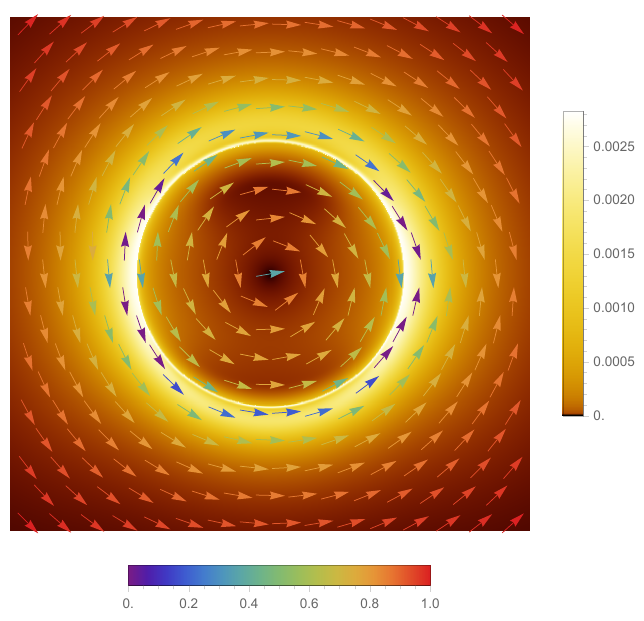}}
\subfigure[$\hat{\lambda}=0.9,\theta_{o}=0.001^\circ$]{\includegraphics[scale=0.4]{Po7_4}}
\subfigure[$\hat{\lambda}=0.9,\theta_{o}=17^\circ$]{\includegraphics[scale=0.4]{Po8_4}}
\subfigure[$\hat{\lambda}=0.9,\theta_{o}=75^\circ$]{\includegraphics[scale=0.4]{Po9_4}}
\caption{Polarized images of black hole shadow in the HOU disk
model. The accretion flow follows the infalling motion, with fixed
parameters $\hat{\gamma}=0.95$ and observing frequency is
$230\,\mathrm{GHz}$.}
    \label{fig13}
\end{figure}

Figures~\textbf{\ref{fig12}} and \textbf{\ref{fig13}} shows that the
impact of the parameters $\hat{\lambda}$ and $\hat{\gamma}$,
respectively, on the polarization images in the HOU disk model. The
bright ring in the images corresponds to the higher-order image, and
the inner dark region originates from the event horizon. All these
results indicates that the polarization intensity $P_{o}$ in the
higher-order image region is significantly stronger than in other
regions, and it rapidly decreases away from this region. The
polarization images exhibit remarkable differences under different
parameter choices, where $\hat{\lambda}$ and $\hat{\gamma}$
determine the intrinsic structure of the space-time, and $\theta_o$
depends on the observers orientation, together they shape the
polarization characteristics. It should be emphasized that in thin
disk models, radiation cannot escape outside the event horizon, so
no polarization effects can be observed in the inner shadow region.
In contrast, in thick disk models, gravitational lensing causes the
horizons outline to be obscured by radiation from outside the
equatorial plane, leading to the presence of polarization vectors
across the entire imaging plane.

\section{Conclusion}
The study of black hole shadow images has recently become a
hot topic in both astrophysics and theoretical physics. The
ground breaking observations of the EHT have opened new avenues for
probing high energy physics and testing gravitational theories,
developing great interest in numerical simulations of black hole
images in the scientific community. During black hole imaging,
the accretion of matter and the associated high energy radiation are
crucial for producing observable features. In this work, we have
considered spherically symmetric static black hole metric in KR
gravity, and investigated the impact of relevant parameters on the
black hole shadow and polarization images. For an unpolarized
images, we have considered two geometrically thick and optically
thin accretion disk models such as, phenomenological model and the
HOU disk model. And for polarized imaging, we only considered the
HOU disk model with anisotropic radiation and accretion flow mode is
the infalling motion. In both cases, we found that the
determination of observable black hole shadow and polarization
images depends on numerous physical parameters inherent to the
accreting flow, including fluid velocity and magnetic field strength
in the fluid's rest frame. In Figs. \textbf{\ref{fig2}} and
\textbf{\ref{fig3}}, we showed that the black hole shadow images of
the phenomenological model with isotropic radiation under fixed
observing frequency at $230 \,\mathrm{GHz}$, and the accretion flow
motion mode is chosen to be the infalling motion. All these images
illustrated that a bright ring-like structure appeared in
the middle of the screen and beyond of this ring-like structure,
there exists a region having non-zero intensity. In the case of
geometrically thick disk models, this region may be obscured by
radiation from outside the equatorial plane, making it difficult to
distinguish. Here, we observed that with the variation of
$\theta_{o}$, as $\theta_{o}$ increases from left to right, the
intensity of higher-order images significantly greater, and the size
of the circular ring decreases with the increasing values of
$\hat{\gamma}$ (see Fig.~\textbf{\ref{fig2}}). The results of Fig.
\textbf{\ref{fig3}} revealed that the increasing values of
$\hat{\lambda}$ reduced the size of the higher-order image, while
increasing $\theta_o$ alter its shape and causes the horizons
outline to be obscured.

Next, we discussed the impact of different observing frequencies and
the accretion flow motion modes in Figs. \textbf{\ref{fig9}} and
\textbf{\ref{fig10}}, respectively. From these results, one can
observe that the increasing values of frequency result in
decrease the corresponding intensity as well as the optical
appearance of the higher-order images and the horizons outline are
clearly visible. Figure \textbf{\ref{fig10}} interpreted that the
participation of the orbiting motion are more prominent in the
combined motion. Moreover, the orbiting motion obscures the horizons
outline with radiation originating from outside the equatorial plane
and notably enhances the overall intensity, while the infalling
motion makes the primary and higher-order images are more distinct.
Consequently, in Figs. \textbf{\ref{fig4}} and \textbf{\ref{fig5}},
we have investigated the impact of $\hat{\lambda}$ and
$\hat{\gamma}$ on the black hole shadow images with an anisotropic
radiation, respectively. These results indicate that the effects of
$\hat{\lambda}$, $\hat{\gamma}$, and $\theta_{o}$ on the black hole
shadow images are similar to those of under isotropic radiation. A
prominent difference is observe that when $\theta_{o}=75^\circ$, the
higher-order image transforms into an elliptical shape under
anisotropic radiation. In Figs. \textbf{\ref{fig6}} and
\textbf{\ref{fig7}}, we discussed the influence of $\hat{\lambda}$
and $\hat{\gamma}$ on the black hole shadow images with anisotropic
radiation, respectively. In all images, there is bright
circular-ring, which corresponds to the higher-order image and
the central dark region comes from the black hole event horizon. The
increasing values of $\hat{\lambda}$ and $\hat{\gamma}$ results in
decrease the size of the circular-ring and the corresponding
brightness increases outside the higher-order image with the
increasing values of $\theta_{o}$. Moreover, the size of the
higher-order image hardly changes, which is in sharp contrast to
the phenomenological model. The influence of different accretion
flow models on the black hole shadow images with different
$\theta_{o}$ are presented in Fig.~\textbf{\ref{fig8}}. These
results showed that at higher observer inclinations, radiation from
outside the equatorial plane more strongly obscure the horizons
outline in the phenomenological model, indicating a stronger
gravitational lensing effect.

Figure \textbf{\ref{fig11}} interpret the numerical results of the
Stokes parameters $\mathcal{I}_{o}, \mathcal{Q}_{o},
\mathcal{U}_{o}, \mathcal{V}_{o}$ on the polarization images in the
HOU disk model with fixed $\hat{\lambda}=0.9$, $\hat{\gamma}=0.55$,
$\theta_o=17^\circ$, and the observing frequency is
$230\,\mathrm{GHz}$. From these images, it is observed that the
parameters $\mathcal{Q}_{o}$, $\mathcal{U}_{o}$, and
$\mathcal{V}_{o}$ reach obtains the peak positions in the region of
the higher-order image and rapidly approaches to zero, as moving
away from this region. In Figs. \textbf{\ref{fig12}} and
\textbf{\ref{fig13}}, we presented the impact of $\hat{\lambda}$ and
$\hat{\gamma}$ on the polarization images in the HOU disk model,
respectively. From these images, one can observe that the
polarization intensity $P_{o}$ in the higher-order image region is
significantly stronger than as compared to other regions, and it
rapidly decreased away from this region. Here the parameter choices
$\hat{\lambda}$ and $\hat{\gamma}$ evaluate the intrinsic structure
of the space-time, and $\theta_o$ depends on the observers
orientation, together they shape the polarization characteristics.
To summarize our results, in the case of thin disk models, radiation
cannot escape beyond the event horizon, so no polarization appears
in the inner shadow. On the other hand, thick-disk models exhibit
gravitational lensing that obscures the horizon with radiation from
outside the equatorial plane, producing polarization vectors across
the entire image. These findings suggest that the characteristics of
black hole shadow as well as polarization images may offer
theoretical evidence for the next-generation EHT observations to
discern the types of different black holes.\\\\
{\bf Acknowledgements}\\
This work is supported by the National Natural Science Foundation of
China (Grants Nos. 12375043, 12575069 ).

\end{document}